\documentclass[preprint,12pt]{elsarticle}

\usepackage{amssymb}
\usepackage{subcaption} 
\usepackage{color}
\usepackage{lineno}
\usepackage{hyperref}
\usepackage{graphicx}
\usepackage{amsmath}
\usepackage{booktabs}

\journal{Nuclear Physics A}

\begin{document}

\begin{frontmatter}



\title{FANSIC: a Fast ANalog SiPM Interface Circuit for the readout of large silicon photomultipliers} 

\author[a]{Luca Giangrande\corref{cor1}}
\cortext[cor1]{Corresponding author: luca.giangrande@unige.ch}

\author[a]{Matthieu Heller} 
\author[a]{Teresa Montaruli} 
\author[a]{Yannick Favre} 

\affiliation[a]{organization={Département de Physique Nucléaire et Corpusculaire, Faculté de Science, University of Geneva},
addressline={24 Quai Ernest-Ansermet},
city={Genève},
postcode={1205}, country={Switzerland}}

\begin{abstract}

Silicon photomultipliers (SiPMs) have increasingly replaced traditional photomultiplier tubes (PMTs) in a wide range of applications, not subject to excessive radiation or extreme temperatures, due to their benefits in terms of compactness, reliability, granularity, lower operating voltage, cost-effectiveness for mass production and photodetection efficiency (PDE).
However,for large-area detection systems, PMTs still remain the preferred choice due to their lower thermal noise and the shorter duration of their output signal, both of which degrade in SiPMs as the sensor size increases. Overcoming these limitations is crucial for the adoption of SiPMs in next-generation gamma-ray observatories.

We present FANSIC, a dedicated analog frontend application specific integrated circuit (ASIC) designed to address these challenges and enable the high-speed readout of large-area SiPMs. Developed in 65~nm CMOS technology, FANSIC features an active summation architecture that efficiently combines signals from multiple SiPM channels and mitigates the undesirable effects associated to with the capacitance of large sensors. Its functional dynamic range extends up to $800$ photoelectrons, while a pulse duration of $3\,$ns and a post-calibration linearity around $5\%$ are achieved within the range of interest, which goes up to $250\,$p.e. These performance are obtained while containing the power consumption to 23~mW per pixel, making FANSIC an ideal choice for experiments in which low-power consumption is critical, like in cameras of large-scale telescopes. 
Designed for versatility, FANSIC offers single-ended and differential output configurations, ensuring compatibility with various digitization solutions and types of SiPMs. Extensive simulations and laboratory measurements confirm its high performance, demonstrating its potential as a key enabling technology for future Cherenkov telescopes.

\end{abstract}



\end{frontmatter}

\section{Introduction}
\label{sec:intro}

FANSIC is an analog ASIC tailored for the conditioning of the electrical signal produced by large SiPM sensors with a surface up to few square centimeters. Its application target is to operate as a pixel front-end for cameras of Imaging Atmospheric Cherenkov Telescopes (IACTs) employing SiPMs, such as for the Advanced Camera of the Large-Sized Telescopes (LSTs) of the Cherenkov Telescope Array Observatory (CTAO)~\cite{Heller:2023qbh} or the SST-1M camera~\cite{SST1Melectronics,Alispach:2024gvp}. The LSTs have a 23\,m diameter segmented mirror reflecting light onto the optical plane with an about 2\,m-size camera collecting it. The SST-1M features a 4-meter-diameter mirror and a camera with a surface area approximately one-fourth that of the LST.

IACTs detect the Cherenkov light produced by extensive air showers that result from the interaction of atmospheric nuclei with gamma and cosmic rays. 
This light is produced in flashes with a typical duration in the range of $5$--$10$\,ns and with wavelengths between $320$\,nm and $550$\,nm (as reported in \cite{CameraPaperHeller2017}) and constitutes the signal of interest. The background is a high rate of continuous light from night sky background (NSB) due to moonlight and its reflections on ground and human activity of tens to hundreds of MHz, electronic and intrinsic sensor noise.

IACTs employ cameras with hundreds to several thousands of pixels in their focal plane. Most of the cameras in operation adopt PMTs because their relatively larger size, compared to SiPMs, allows to cover square-meter surfaces with fewer channels and less complexity~\cite{ray2007minibooneexperimentoverview, okubo2023mcp, akindele2023acceptance, allekotte2008surface}. 
However, since 2010 the ETH in Zürich and the University of Geneva have pioneered the adoption of SiPMs in IACTs cameras through the development of the FACT telescope at La Palma, Canary Islands, Spain~\cite{FACT} and of the SST-1M telescopes~\cite{Alispach:2024gvp} in operation at the Ond\v{r}ejov observatory in the Czech Republic~\cite{CameraPaperHeller2017}. The SST-1M is equipped with a camera with $1296$ pixels arranged in a hexagonal configuration with a long diagonal of approximately $1$\,m. 
This telescope was conceived to provide the 37 Small-Sized Telescopes (SSTs) of CTAO, the new generation of gamma-ray observatory targeting energies in the range $20\,$GeV--$300$\,TeV by combining small, medium (MSTs) and Large-Sized telescopes. As a matter of fact, the energy threshold of telescopes depends on the inverse of the linear size of the mirrors, so that the LSTs cover the lower energy region and the SSTs dominate the CTAO sensitivity beyond 10\,TeV. The SSTs of CTAO will be dual-mirror telescopes, unlike SST-1M and employ SiPMs as well.

The acquisition of the physical signal is accomplished by means of transduction to the electrical domain via photomultiplication, conditioning and conversion into numerical data by the front-end electronics. The term frontend is used to identify the electronics responsible for the signal acquisition in contrast with backend which applies to the post-processing of data such as events discrimination and triggering.

FANSIC was designed to satisfy the requirements of the camera upgrade planned for LSTs and discussed in~\cite{Heller:2023qbh}. 
This camera has a hexagonal shape as the SST-1M camera and about two times its linear size.

In the last decade, several works have investigated the development of SiPM modules for IACT cameras. The challenge of the application of SiPMs over large sensitive areas is achieving a sub-nanosecond resolution for at least 10\,p.e. and at the same time keeping the signal as short as possible for fast bandwidth requirements due to high rates of signal and to avoid integrating during signal duration  large NSB. Additionally, it is important to contain power budget. To this end, several analog frontends have been developed adopting both commercial components~\cite{DEPAOLI2023168521, HAHN2024169350} and existing ASICs~\cite{De_Angelis_2023, ARCARO201726}.
The main characteristics of FANSIC are reported in Tab.~\ref{tab:FANSic_specs}. The chip is a prototype designed to verify that the performance obtained with a standard CMOS $65$\,nm technology node satisfies the application requirements. The main goals are to guarantee a pulse duration of $3\,$ns and noise levels compatible with single-photoelectron resolution, while containing the power consumption. In addition, the compatibility with an input capacitance up to $1\,$nF makes FANSIC viable for applications that target temporal resolutions in the nanosecond range together with pixel areas up to $\sim1\,\text{cm}^2$. This is the case for many experiments that employ large array of photodetectors to cover surfaces of several square-meters. 
\begin{table}
\centering
\resizebox{11cm}{!}{
\begin{tabular}{llr}
\toprule
\midrule
Technology          & CMOS $65$\,nm \\
Dimensions          & $3.584\times3.584$\,mm$^2$ ($12.845$\,mm$^2$ )  \\
Inputs              & $32$ inputs \\
Outputs             & $8$ outputs \\                         
Power supply        & $1.2$\,V \\
Power consumption   & $184$\,mW \\
Adjustable features  & Gain, pulse shaping and single-ended baseline \\
Package             & Bare dies \\ 
\midrule
\bottomrule
\end{tabular}
}
\caption{Main characteristics of the FANSIC chip}
\label{tab:FANSic_specs}
\end{table}

The frontend requirements are detailed in Section \ref{design_req} of this manuscript. Section \ref{frontend_design} deals with the considerations pertaining the design of the main circuits within the ASIC, whereas Section \ref{experimental_results} illustrates the characterization procedure together with the most relevant measurement results. Finally, in Section \ref{conclusions} some conclusions are drawn.

\section{Design requirements} \label{design_req}
This section illustrates the application requirements that drove the design of the analog frontend which will be installed just behind the photodetection plane of the camera, to minimize the distance from the photodetectors.\\
\noindent
\subsection{Sensor}
The upgraded camera will replace PMTs with SiPMs to improve the telescope's sensitivity and spatial resolution. Currently, the pixels consist of round PMTs, with a diameter of $3.81$\,cm, coupled to hexagonal light guides. Whereas, the upgraded camera will feature hexagonal SiPMs with a long diagonal below $1$\,cm, allowing smaller light guides and optimal use of the photosensitive area.
SiPMs are built as arrays of square shaped single-photon avalanche photodiodes (SPADs) with a size in the tens of micro-meters. This value drives the trade-off of the main indicators of the performance of these photodetector, such as the PDE, dark count rate (DCR), optical cross-talk, dynamic range, and time resolution. For applications like the cameras of IACTs, the optimum is found around $50$--$75\,\mu$m, resulting in several thousands of micro-cells per pixel. Such dimensions maximize the PDE. 

The LSTs are installed at Los Roque de los Muchachos at La Palma island. This location was selected for the construction of CTAO Northern array, which will feature a total of 13 telescopes and will be completed in the coming years.
At this specific site, when observing galactic sources, each pixel of the LST camera detects an average NSB rate around $300\,$MHz. This is the primary background for the signal of the Cherenkov light produced by extensive air showers. The NSB and cosmic ray or photon images can be distinguished by their spectral and temporal characteristics~\cite{CameraPaperHeller2017}.

The Cherenkov signal appears with sporadic flashes, usually lasting from a few to $20\,$ns, and is spatially confined within a cluster of pixels. In contrast, NSB photons arrive continuously, exhibiting a uniform statistical distribution in both space and time.

A SiPM prototype intended for the camera upgrade is being developed in collaboration with Hamamatsu leading to the series S13360, which provides enhanced sensitivity in the near-UV spectrum, faster response, and single-photon resolution. FANSIC has been simulated and tested with a $3\times3\,$mm$^2$ S13360-3075CS-UVE SiPM, referred to as the S13360 SiPM for the remainder of this manuscript. 

\subsection{Active Summation}
Compared to the PMTs currently in use, the proposed SiPMs demonstrate nearly double the sensitivity in the wavelength range of $300\text{--}600\,$nm~\cite{miranda}. This increased sensitivity allows for a significant reduction in pixel area while maintaining a comparable number of photoelectrons per readout channel.

The required pixel area can be achieved through custom hexagonal designs or by assembling commercially available SiPMs, such as combining four monolithic $6\times6$\,mm$^2$ SiPMs into a single tile with a shared cathode. However, this relatively large pixel area presents notable challenges for frontend design, particularly in terms of noise, power consumption, and speed. To address these challenges, the adopted strategy involves partitioning the pixel into four sections. Each section is read out independently while meeting the required performance specifications. The signals from the four sections are then merged into a single readout channel, reducing the number of lines in the camera while ensuring optimal performance.

Hence, the ASIC has to feature an active summation of multiple input channels as done in~\cite{MUSIC}. 
The size and number of the pixel sub-sections are mostly constrained by the resolution requirements. A major limitation comes from the thermal noise charges stored in the capacitors of the SiPM's electrical model, such as those presented in \cite{corsi2007modelling, marano2013silicon, marano2014accurate, acerbi2019understanding}. 
These contributions can be grouped into a single noise charge associated with the equivalent output capacitance of the SiPM, which is given by:
\begin{equation}
    C_{\text{eq}} = \frac{C_q \cdot C_d}{C_q + C_d} N + C_g
    \label{eq:capacitor_combination}
\end{equation}
where, adopting the same naming convention as \cite{corsi2007modelling}, $C_q$ is the quenching capacitor, $C_d$ is the diode capacitance and $C_g$ is the parasitic capacitance dominated by the contribution of the conducting grid among the SPADs.
The noise charge can then be converted into an equivalent number of photoelectrons (ENP) by dividing it for the charge produced by one photoelectron, according to:
\begin{equation}
ENP = \frac{\sqrt{k_b \cdot T\cdot C_{eq}}}{|q_e\cdot G|}
\label{eq:ENP}
\end{equation}
where $k_b$ is the Boltzmann constant, $T$ is the temperature in Kelvin, $q_e$ is the electron charge, $G$ is the SiPM gain and $C_{eq}$ is the equivalent output capacitance of the SiPM specified in Eq.~\ref{eq:capacitor_combination}.
Using the estimated values of $C_{eq}=650$\,pF, $G=10^4$ and $T=290$\,K, the equivalent noise reaches almost one photoelectron. To ensure adequate noise levels to achieve single photoelectron resolution, the pixel is divided into four equal sections, reducing the ENP below $0.5$.

\subsection{Speed}
\begin{figure}
    \centering
    \includegraphics[width=0.7\linewidth]{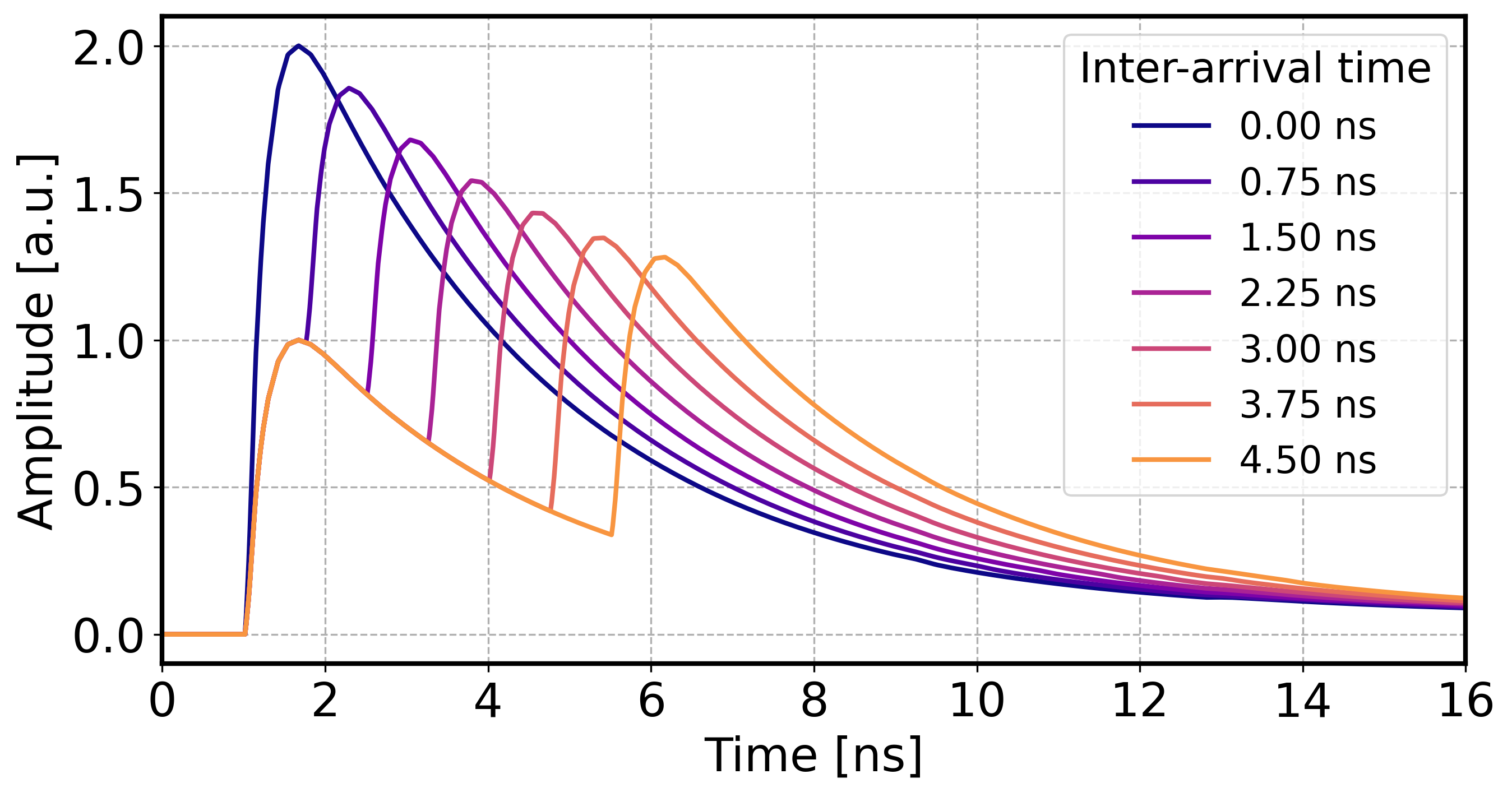}
    \caption{Numerical simulation of a pixel output signal for various inter-arrival times of two photoelectrons with a FWHM of $3\,$ns. The amplitude is normalized to the peak value of a single photoelectron. With a threshold on the amplitude of $1.5$, the coincidence window extends up to $2.5\,$ns.}
    \label{fig:coincidence}
\end{figure}

The avalanche charge is converted into voltage through a resistor connecting the anode of the SiPM to the ground potential. The resulting signal is a positive pulse with three distinct time constants, which are included in the electrical models of the SiPM. The rising $\tau_\mathrm{r}$ and the two falling time constants, a faster $\tau_\mathrm{ff}$ and a slower one $\tau_\mathrm{fs}$, can be calculated as:
\begin{equation}
\begin{cases} 
    \tau_\mathrm{r} = R_\mathrm{d} \cdot C_\mathrm{d} \\
    \tau_\mathrm{ff} = R_\mathrm{s} \cdot C_\mathrm{eq} \\
    \tau_\mathrm{fs} = R_\mathrm{q} \cdot (C_\mathrm{q} + C_\mathrm{d})
\end{cases}
\label{eq:taus}
\end{equation}
where $R_\mathrm{d}$ is the diode resistance, $R_\mathrm{q}$ is the quenching resistance and $R_\mathrm{s}$ is the external resistance to ground. For a S13360 SiPM with a series resistor of $50\,\Omega$, the values to the time constants are approximately $100\,$ps for $\tau_\mathrm{r}$, $1$\,ns for $\tau_\mathrm{ff}$ and $1\,\mu$s for $\tau_\mathrm{fs}$. 
The large separation between the two falling time constants makes the smaller one dominant for the initial evolution relevant for the pulse duration which is measured as full width at the half maximum value (FWHM). Whereas the larger one introduces a slowly decaying tail which is discussed in detailed in the next Section.
The pulse FWHM affects the requirements of the electronics used to acquire and the signal and those of the trigger system. 
Fig.~\ref{fig:coincidence} illustrates the simulation of the electrical signal of a pixel induced by two photoelectrons in case of a FWHM of $3\,$ns. 
This value allows limiting the acquisition sampling rate to $1\,$GHz, thus relaxing the bandwidth and power requirements, and provides a reasonable lower bound for the threshold.

\begin{figure}
    \centering
    \includegraphics[width=0.7\linewidth]{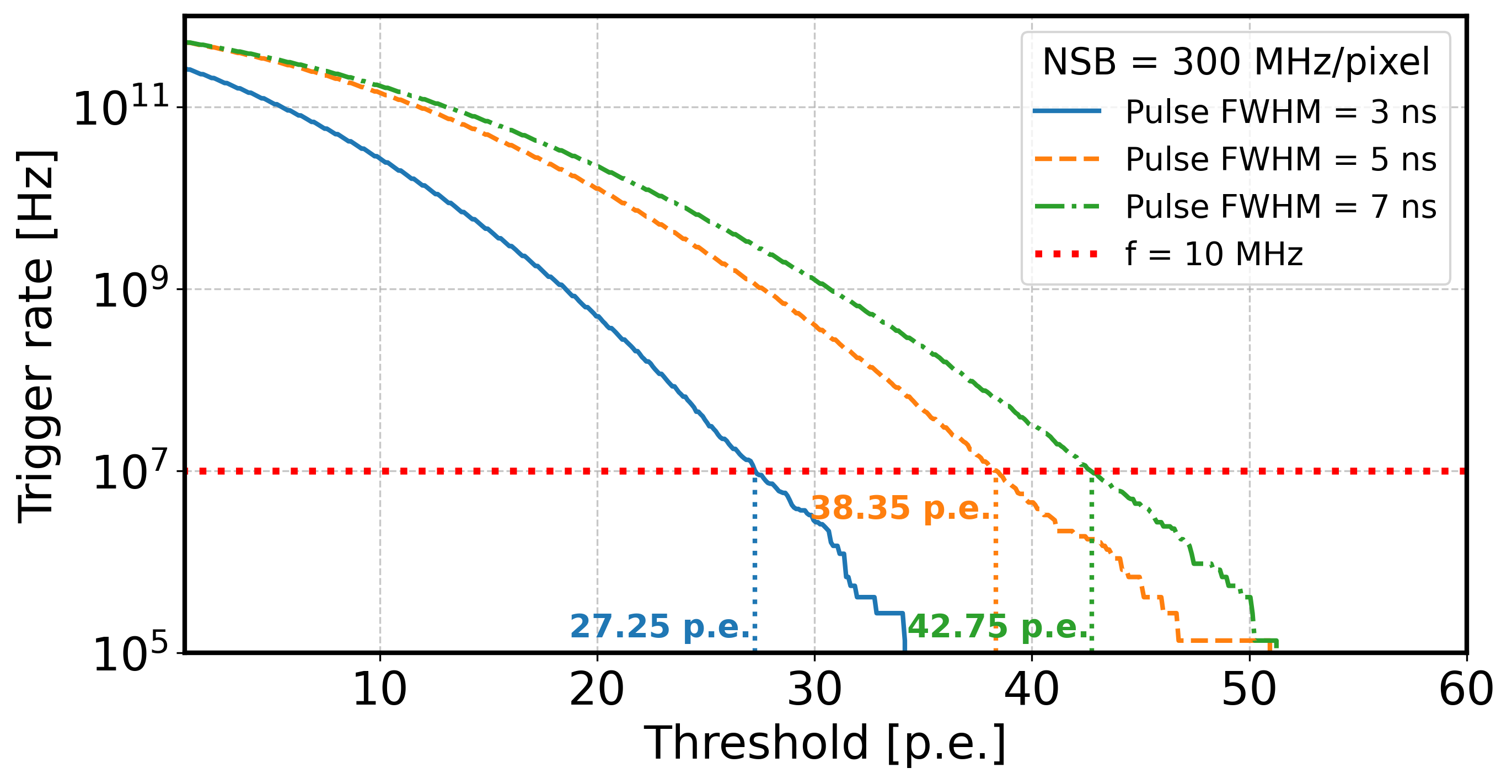}
    \caption{Simulation of the trigger system under NSB conditions. The vertical axis shows the average triggering rate as a function of the threshold, set in number of photoelectrons, for various pulse widths. The camera is organized in $1141$ clusters, each with $49$ pixels. Each cluster is monitored at a rate of $1\,$GHz and the triggering condition is met whenever the sum over a cluster exceeds the threshold, set as number of photoelectrons.}
    \label{fig:trig_rate}
\end{figure}

In fact, Fig.~\ref{fig:trig_rate} illustrates the effect of the pulse FWHM on the triggering capabilities of a Cherenkov camera. The numerical simulation was performed according to the design described in \cite{Heller:2023qbh}. The maximum trigger rate that the system can handle is $10\,$MHz, therefore the FWHM of the pulses determines the lower bound for the threshold. Longer pulses lead to higher probabilities for NSB photoelectrons to overlap and trigger a camera readout. For instance, extending the FWHM from 3 to $5\,$ns would imply a $40\%$ increase of the threshold, thus significantly affecting the scientific reach of the instrument.

The FWHM of the SiPM's output pulse is about $2$\,ns, thus the electronics needs to shape the pulse in order to obtain the desired duration. This translates into a bandwidth requirement for the amplifiers which is discussed in Section \ref{frontend_design}. 

\subsection{Dynamic range}
\begin{figure}
    \centering
    \includegraphics[width=0.7\linewidth]{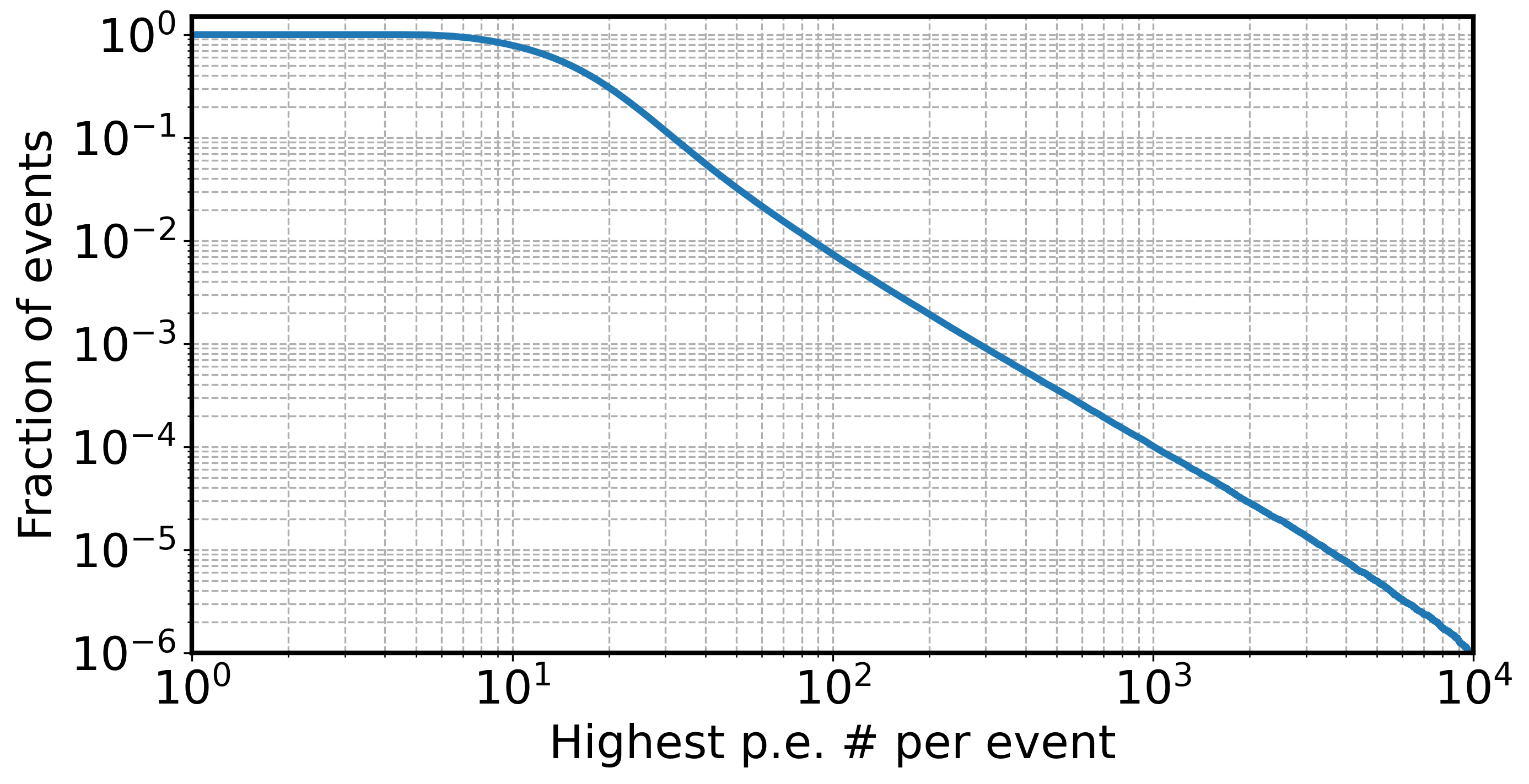}
    \caption{Simulation a Crab-like spectrum~\cite{ALEKSIC201530} of high-energy gamma events captured by the camera. The simulation includes the models of the sensor and electronics reported in this work. The graph shows the fraction of events, relative to the total, that contain the maximum number of p.e. reported on the horizontal axis.}
    \label{fig:max_npe_pixel}
\end{figure}

The dynamic range requirement is introduced in order to maximize the telescope's capability of capturing the targeted astrophysical events. For the bulk of the shower simulations depicted in Fig.~\ref{fig:max_npe_pixel}, the fraction of events with a single pixel collecting more than $100\,$p.e. is less than $1\%$ and more than $250\,$p.e. is less than $0.1\%$. Therefore, the dynamic range in which ASIC performance need to meet the requirements hereby listed is set from $1$ up to $250\,$p.e. 
Beyond this range, the FWHM requirement does not apply and a degradation on the charge resolution is expected, as shown in figure~47 of \cite{CameraPaperHeller2017}. 
To prevent hardware damage, it is crucial that the frontend allows the identification of out-of-range light intensities. This capability ensures that excessive power dissipation, which could lead to hardware failures, mitigated by triggering a fail-safe mechanism.

The total capacitance observed at the anode of a SiPM depends on the sensor technology and design, with typical values reaching hundreds of picofarad for a surface around $1\,\text{cm}^2$ \cite{Aguilar:2016iit, acerbi2019understanding, Heller:2023qbh, corsi2007modelling}. This relatively large capacitance is dominated by the parasitic and quenching capacitors of the electrical model, as described in Eq. \ref{eq:capacitor_combination}. It introduces a noise contribution detailed in Eq.~\ref{eq:ENP} and an attenuation of the SiPM signal which have a negative effect on the dynamic range. The attenuation is related to the fact that the avalanche charge finds an additional low-impedance path to ground through this capacitance thus leading to a reduced voltage across the external resistor.

\subsection{Signal to noise ratio}
The requirement for the signal-to-noise ratio (SNR) originates from the need to detect single photoelectron events with a false-positive incidence smaller than $1\%$. with a threshold of half a photoelectron. Defining the SNR as the ratio of the mean over the standard deviation of a normal distribution allows to quantify the requirement based on the false-positive probability. Having an SNR of 5 in case of $1\,$p.e. means that a discrimination threshold placed at half a photoelectron will be 2.5 standard deviations away from both the expectations of $0$ and $1\,$p.e., resulting in a false-positive probability of:
\begin{equation}
    P_\mathrm{false-pos} = \frac{1}{\sqrt{2\pi}}\int_{2.5\sigma}^{\infty} e^{-x^2/\sigma^2}\cdot dx\approx0.62\%
    \label{eq:false_pos}
\end{equation}
The main noise contributions come from the SiPM and the frontend electronics. The separation of the two allows for the evaluation of the ASIC performance independently of the sensor used for its characterization. 
Since the contributions are independent from each other, the electronic SNR is simply defined as:
\begin{equation}
   \mathit{SNR}_\mathrm{e} = \frac{G(N_\mathrm{pe}=1)}{\sigma_\mathrm{e}} 
    \label{eq:snr}
\end{equation}
where $G(N_\mathrm{pe}=1)$ is the detector gain for the first photoelectron and $\sigma_\mathrm{e}$ is the contribution of the electronics to the standard deviation of the normal distribution.

\subsection{Output loads}
The analog signal provided by the frontend ASIC needs to be digitized in order to enable the large amount of computations required to identify the different types of air showers. 
The digitization involves acquiring the signal with an analog-to-digital converter (ADC), which in this user case samples at a fixed rate of $1\,$GHz, and estimating the SiPM charge by reconstructing the area under the pulse.
To this end, an integrated ADC, either a commercial such as the ADC12xJ1600 series from Texas Instruments, or a custom one will be employed. In order to accommodate ADC architectures with single-ended and differential inputs, the ASIC needs to be able to drive both single-ended and differential loads without requiring additional components.

\subsection{Linearity}
The linearity within the specified dynamic range must be better than $5\%$ after calibration. This specification derives from a higher-level requirement from CTAO: \textit{The average pixel response for input signals within the relevant detection range must be known to better than $8\%\,$rms at all times}.
The main contributors to the non-linearity of the pixel response are the SiPM, the frontend electronics, the optics and the digital electronics. A requirement of $5\%$ was assigned to the frontend to ensure sufficient margin in the global error budget for the other contributions.
 
\section{Frontend design} \label{frontend_design}
\begin{figure}[h!]
    \centering
    \includegraphics[width=0.9\textwidth]{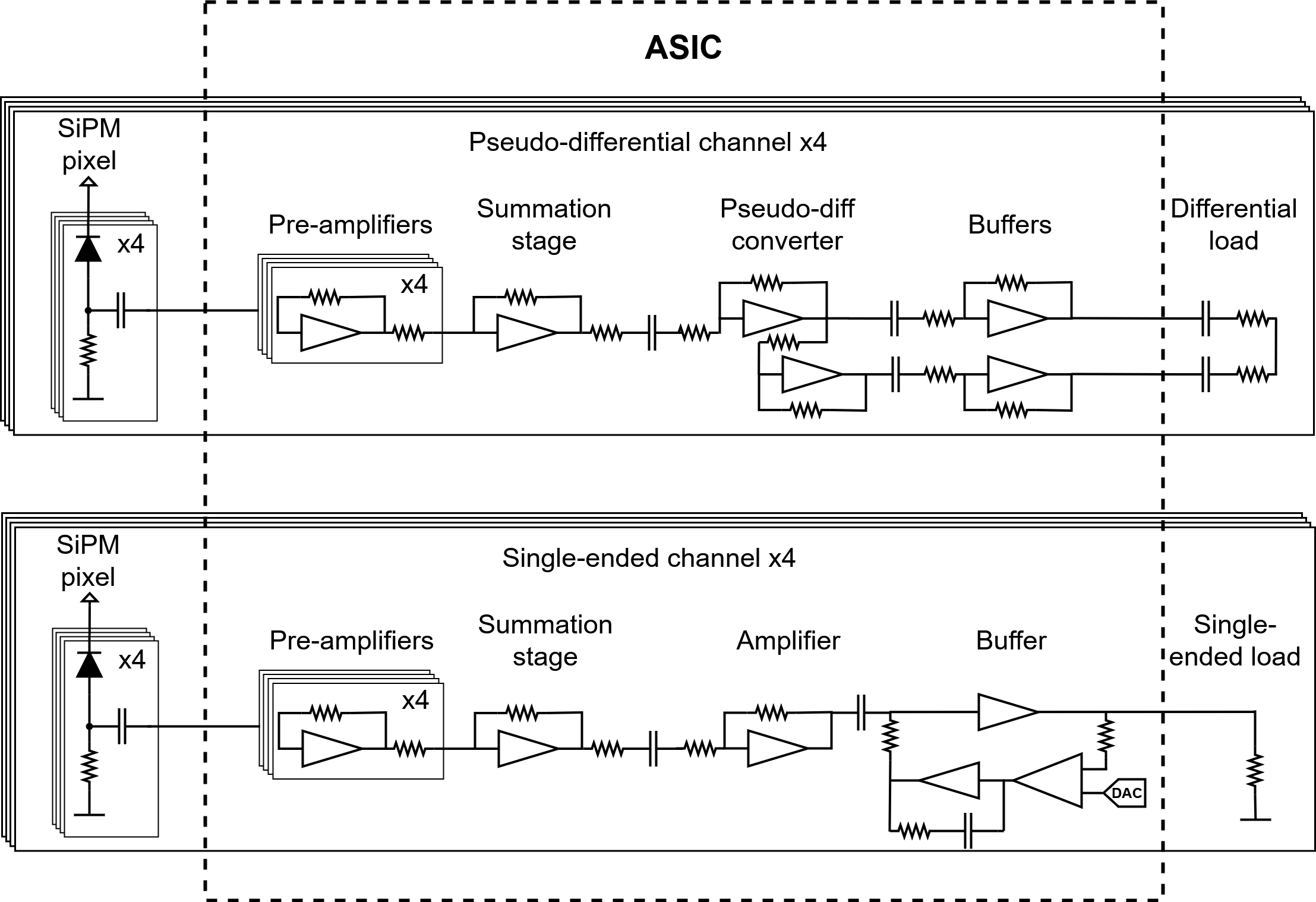}
    \caption{Simplified FANSIC block diagram. The ASIC contains four  pseudo-differential front-end channels and four single-ended ones. Each channel processes the signal of an entire pixel and is built as a succession of four signal processing stages. Altogether the ASIC features 32 inputs and 8 outputs. }
    \label{fig:block_level}
\end{figure} 

The architecture chosen for the development of FANSIC is shown in Fig.~\ref{fig:block_level}. The chip contains $8$ channels, each conditioning the signal of one pixel provided through $4$ independent input lines. In order to comply with the output load requirements and assess the influence of electromagnetic interference within the camera, two types of buffer stages have been designed. 
The purpose of buffering is to preserve the characteristics of the voltage pulse feeding the external ADC through transmission lines. Both buffers are compatible with $50\,\Omega$ matching and are therefore capable of driving the lines termination resistors. 
Each buffer design is included within a dedicated channel type. These are therefore distinguished on the basis of the supported output configuration, either the pseudo-differential or single-ended one. 

The pseudo-differential channel comprises an input stage made of four parallel pre-amplifiers, an active summation stage, a single-ended to pseudo-differential converter and an output stage that buffers the signal towards a (pseudo-)differential resistive load. 

The single-ended channel contains the same input and summation stages but an additional amplification stage is used in place of the single-ended to pseudo-differential converter, and a different topology is adopted for the output stage. 
Specifically, this output stage features an integrated 5-bit digital-to-analog converter (DAC), which provides the reference voltage for the regulation of the output DC voltage. The DAC architecture employs one resistor and a collection of binary scaled current mirrors to produce an output voltage that covers the range $20\text{--}190\,$mV. This versatility, together with the gain adjustable features described in subsection 3.4,
allows an optimal input range occupation of the external ADC.   
In fact, the channels feature adjustable bias currents for the amplifiers and adjustable AC-coupling capacitors to accommodate different types of sensors and optimize the output pulse characteristics.
The single-ended buffer is designed to be DC-coupled to its load whereas the pseudo-differential one to be AC-coupled in order to prevent unnecessary restrictions on the input common mode of the external ADC.

From the electrical point of view, a SiPM behaves as an inversely polarized diode that outputs a charge in response to photoelectrons. The cathode is biased with $56$\,V for the S13360 whereas the anode is connected to ground through a series resistor. The resistor also transduces the SiPM signal into a voltage through Ohm's law. This signal is fed to the front-end ASIC via an external AC-coupling capacitor which allows the independent biasing of the SiPM anode and the input terminal of the pre-amplifiers.

\paragraph{\bf Signal conditioning}
The signal produced by the S13360 SiPM is a current pulse with a sharp rising time, of the order of hundreds of ps, and two decaying time constants: a faster one of around $1$\,ns and a slower one of around $1$\,µs, whose derivation from the electrical model was given in Eq. \ref{eq:taus}.

\begin{figure}[h!]
    \centering
\includegraphics[width=0.7\textwidth]{ 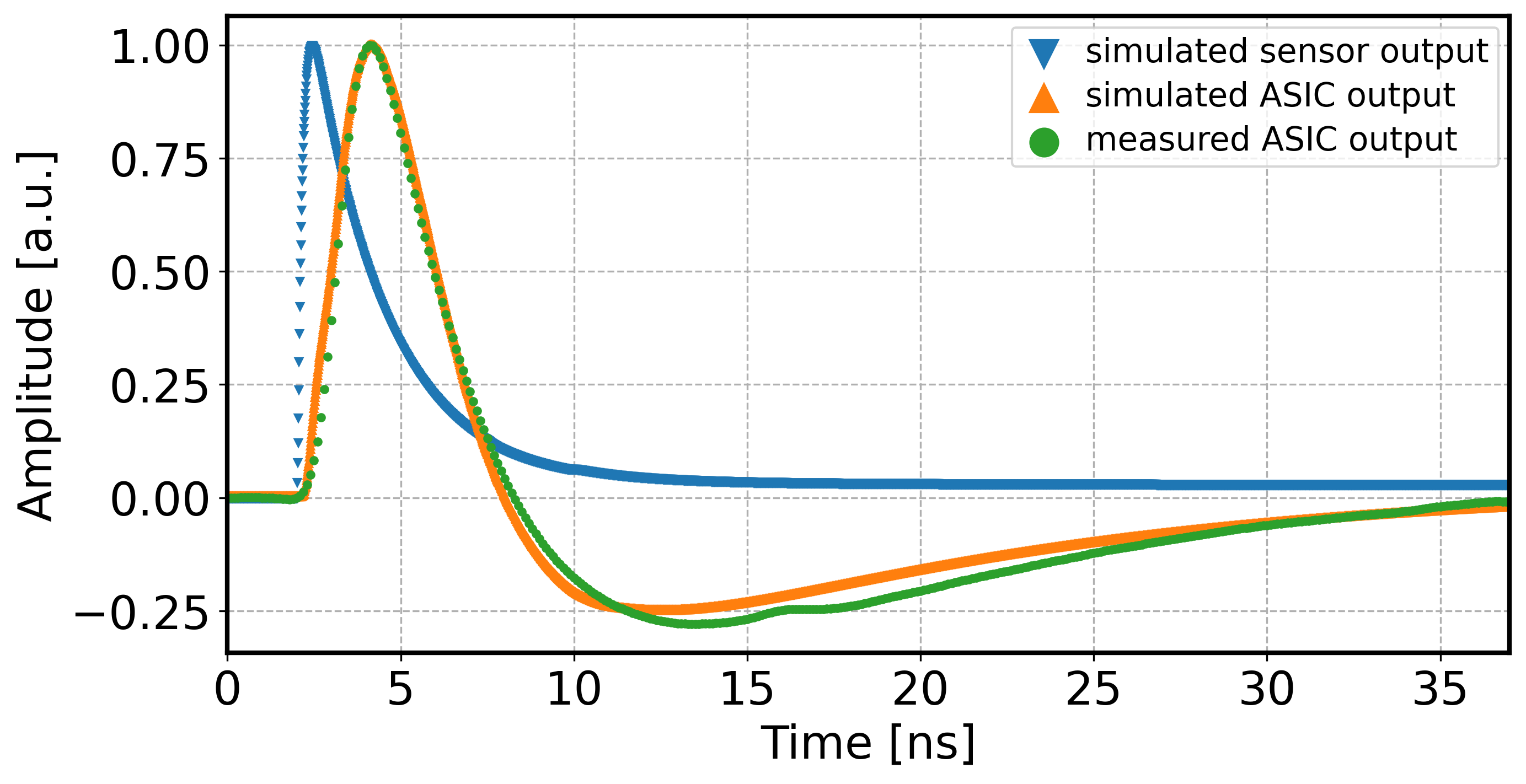}
    \caption{Pulse shape comparison between the simulated S13360 current (blue) and the simulated (orange) and measured (green) frontend output voltages with normalized peak amplitudes. The measured output corresponds to the average of ten thousand waveforms acquired under fixed light conditions with an expectation of $5\,$p.e.}
    \label{fig:in_out_pulses}
\end{figure} 

\begin{table}[h!]
\resizebox{\textwidth}{!}{\begin{tabular}{lcccc}
\toprule
\midrule
 & Unit & SiPM simulation & FANSIC simulation & FANSIC measurement \\
Rising time& [ns] & $0.23$ & $1.25$ & $1.13$ \\
FWHM &[ns] & $1.99$ & $3.05$ & $2.82$ \\
Peak-to-peak &[ns] & $-$ & $8.44$ & $9.29$ \\
$10\%$ recovery time &[ns] & $5.67$ & $20.74$ & $21.61$ \\
$1\%$ recovery time &[ns] & $1249$ & $95.44$ & $30.59$ \\
$0.1\%$ recovery time &[ns] & $3752$ & $808.1$ & $33.08$ \\
\midrule
\bottomrule
\end{tabular}}
\caption{Original data points with a time step of $100\,$ps interpolated with a cubic method and a time step of $5\,$ps. The recovery times describe the baseline recovery and correspond to the time interval from the peak to the crossing of a threshold relative to the peak. }
\label{tab:pulses}
\end{table}

The main effects of the signal conditioning performed by FANSIC is shown in Fig.~\ref{fig:in_out_pulses}. The pulse duration is slightly increased and the slowly decaying tail is replaced by an undershoot with shorter recovery, detailed in Tab.~\ref{tab:pulses}.
The time constant $\tau_\mathrm{fs}$ is associated with the quenching mechanism which extinguishes the avalanche by reducing the internal electric field of the SiPM. 

From an electrical standpoint, however, the long exponential decay of the pixel signal is undesirable. It causes a significant gain attenuation and performance degradation, ultimately reducing the resolution for observation of events.

The pixel gain attenuation is due to the behavior of its micro-cells, which cannot reproduce the nominal response until the recharge process is complete. 
The degree of attenuation depends on the value of $\tau_\mathrm{fs}$, the NSB rate and the area of the pixel.
The time evolution of a micro-cell's charge following an avalanche event is given by: 
\begin{equation}
    q(t) = q_0 \cdot \left(1 - e^{-\frac{t}{\tau_\mathrm{fs}}}\right)
    \label{eq:charge}
\end{equation}
where $q_0$ represents the steady state charge. This equation implies that a complete recharge can only be achieved asymptotically. Nonetheless, in practice, a recharge can be considered completed after few time-constants (for instance, an interval of $4.7\,\tau_\mathrm{fs}$ is sufficient to reach $0.99\cdot q_0$). 
The dependence on the NSB rate is due to the fact that the vast majority of detected photons belong to the background. Under the assumption of random uncorrelated occurrences with a constant average rate, which is satisfied in the usual observation conditions, the time between consecutive detections per micro-cell can be modeled with the exponential distribution:  

\begin{equation}
     \mathcal{P}(r,t) = r \frac{A_\mathrm{cell}}{A_\mathrm{pix}}\cdot  e^{-r \frac{A_\mathrm{cell}}{A_\mathrm{pix}}\cdot t}
    \label{eq:nsb_dist}
\end{equation}
where the photoelectrons NSB rate per micro-cell is determined by scaling the pixel rate $r$ according to the ratio of the respective areas, $A_\mathrm{cell}$ and $A_\mathrm{pix}$.
The expectation value of the micro-cell charge can then be calculated taking into account the probability that a given time has passed since the last avalanche, according to:

\begin{equation}
    \bar{q}(r) = q_0 \cdot \left(1-\int_{t_0}^{\infty}  \mathcal{P}(r,t)e^{-t/\tau_\mathrm{fs}} \cdot dt\right) \approx \frac{q_0}{1+{r\tau_\mathrm{fs}\cdot \frac{A_\mathrm{cell}}{A_\mathrm{pix}} }}
    \label{eq:gain_nsb}
\end{equation}
where $t_0$ represents the micro-cell dead time. The approximation relies on the fact that the dead time is several orders of magnitude smaller than the recharge time constant, thus providing a negligible contribution.

\begin{figure}
    \centering
    \includegraphics[width=0.7\textwidth]{ 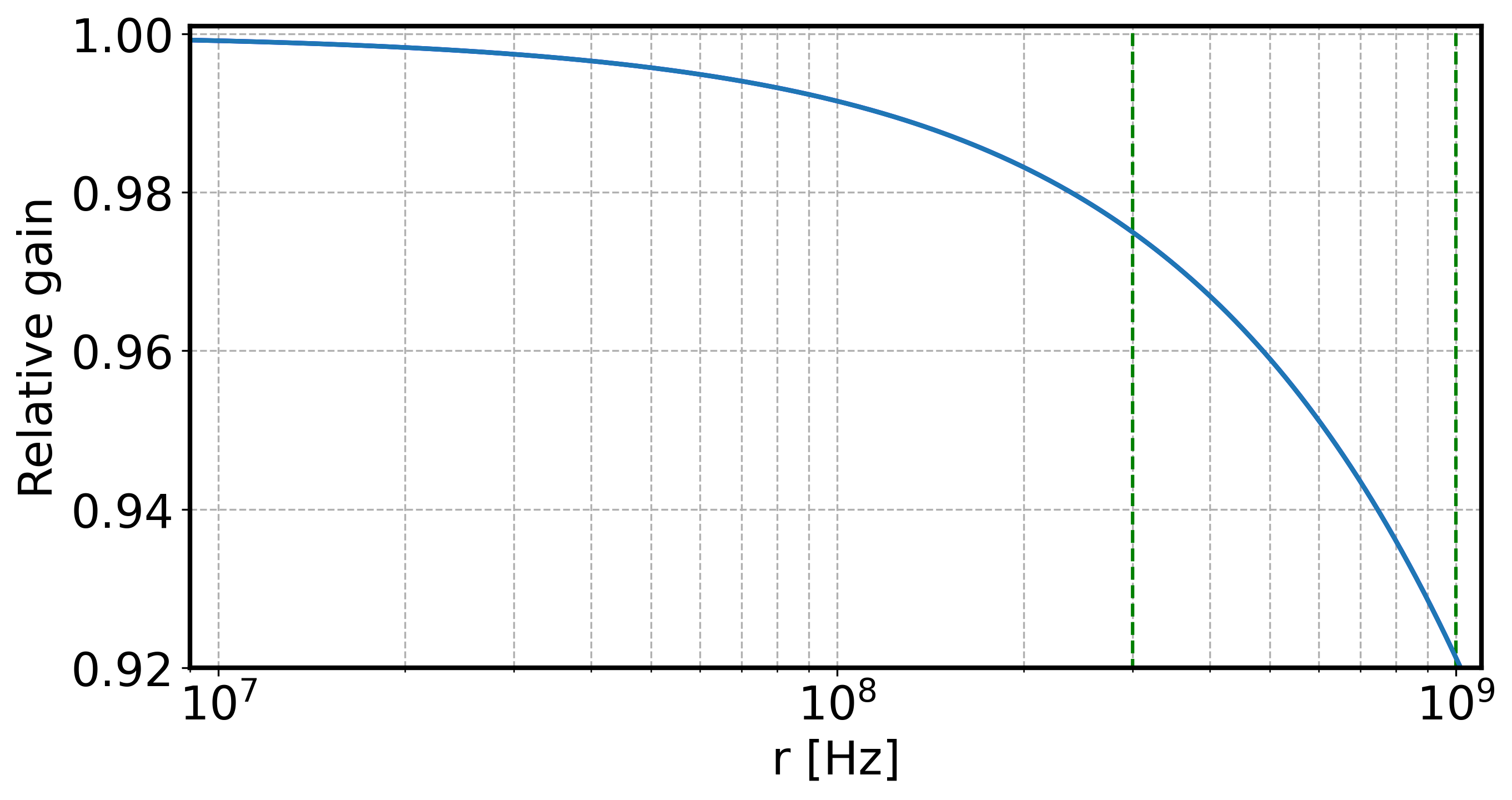}
    \caption{Simulation of the effective gain of a hexagonal S13360-UVE pixel as a function of the NSB rate per pixel. The gain is normalized to its nominal value which is unaffected by the NSB. The vertical dashed lines correspond to the nominal ($300$\,MHz) and maximum ($1$\,GHz) rates expected under operational conditions.}
    \label{fig:pix_gain}
\end{figure}

The dependence of the pixel gain on the rate $r$ is numerically computed, using the charge expectation (Eq.~\ref{eq:gain_nsb}) rather than its nominal value $q_\mathrm{0}$ and the result is shown in Fig.~\ref{fig:pix_gain}. As expected, higher rates correspond to larger gain losses due to the shorter mean time between consecutive micro-cell avalanches. 
The typical and maximum NSB photoelectron rates per pixel of $300$\,MHz and $1$\,GHz depicted in the figure correspond to the values observed with the PMT camera currently equipped by LST. The same values are used as a reference for the design of the SiPM camera because the difference in size between the two types of pixels is expected to be compensated by the difference in sensitivity. The gain attenuation induced by the NSB is an intrinsic property of the SiPM which is directly related to its design parameters such as the quenching resistance and the number of micro-cells per pixel.

\begin{figure}
    \centering
    \includegraphics[width=0.7\textwidth]{ 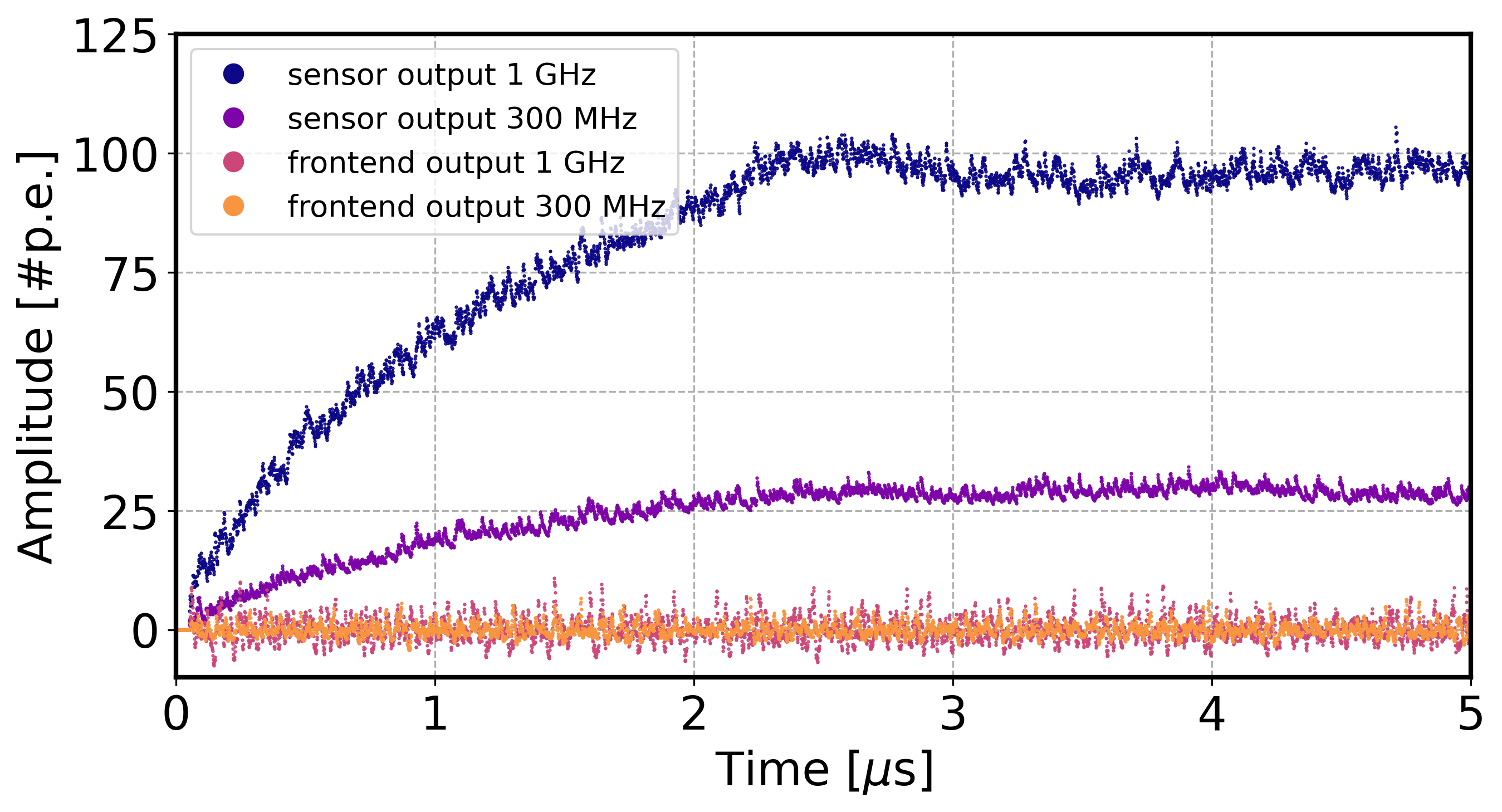}
    \caption{Simulation of the NSB pile-up time evolution. A Poisson distribution of photons is injected at $t=0$\,s with rates of $300$\,MHz and $1$\,GHz. The amplitudes are shown in units of p.e. The baseline shift observable at the output of the SiPM settles within $\sim2\,\tau_\mathrm{fs}$ and is properly filtered by the ASIC.}
    \label{fig:pile-up}
\end{figure}
The degradation of the pixel dynamic range is a consequence of the pile-up of the slowly decaying pulse tails, which result in a baseline shift proportional to the NSB rate. 
Unlike the gain attenuation, this undesirable effect can be prevented through proper filtering of the electrical signal going through the frontend.

Since the time constants $\tau_\mathrm{ff}$ and $\tau_\mathrm{fs}$ differ by three orders of magnitude, it is possible to perform the filtering without affecting the fast component of the pulse. The optimal filtering characteristic is implemented through the introduction of properly sized AC-coupling capacitors throughout the channels, visible in the block-level schematic. 
A circuit simulation is performed to confirm the filter efficacy for the expected and maximum NSB rates, and the results are reported in Fig.~\ref{fig:pile-up}. In both cases, the baseline shift observable at the SiPM output is successfully removed throughout the frontend and the entire dynamic range is available for detection of atmospheric showers. 
In addition, the AC-coupling provides the design flexibility of defining the bias point of the various stages independently from each other.

\paragraph{\bf Bandwidth considerations.}
The largest portion of the charge generated within a SiPM is contained in the area under the slow component of the pulse, associated with $\tau_\mathrm{fs}$. 
In order to enable fast detection and prevent the aforementioned loss of dynamic range, only the fast component of the pulse is used by limiting the region of interest to a few $\tau_\mathrm{ff}$.
This implies that just a few percentiles of the avalanche charge and intrinsic noise charge of the SiPM are relevant for detection.
The overall transfer function of the ASIC channels shows a band-pass filtering characteristic, where the high-pass part removes the tails and the low-pass part filters the excess Johnson-Nyquist noise and contributes to the final pulse shape.
For a S13360 SiPM, about 1\% of the charge is delivered in $2\tau_\mathrm{ff}$ and its pulse can be shaped with a high-pass frequency up to $30\,$MHz and a low-pass frequency down to $250\,$MHz while preserving the desired features.

The transfer function that filters the pulse and determines its shape contains a low-pass and a high-pass component. The former is responsible for the extended pulse duration, a rising time increase of about $5\times$ and the introduction of a delay around $1\,$ns. The latter speeds up the recovery of the ASIC output pulse compared to the SiPM, thus preventing the aforementioned baseline shift. This feature comes at the cost of a short-lived undershoot. 
A detailed comparison of the pulse characteristics between the outputs of the SiPM and FANSIC (simulated and measured) is reported in Tab.~\ref{tab:pulses}. 

The channel transfer functions in FANSIC can be adjusted through digital settings to allow for optimization of the pulse shaping characteristic and to accommodate different SiPMs. 

\subsection{Input stage}
\begin{figure}
    \centering
    \includegraphics[width=0.6\textwidth]{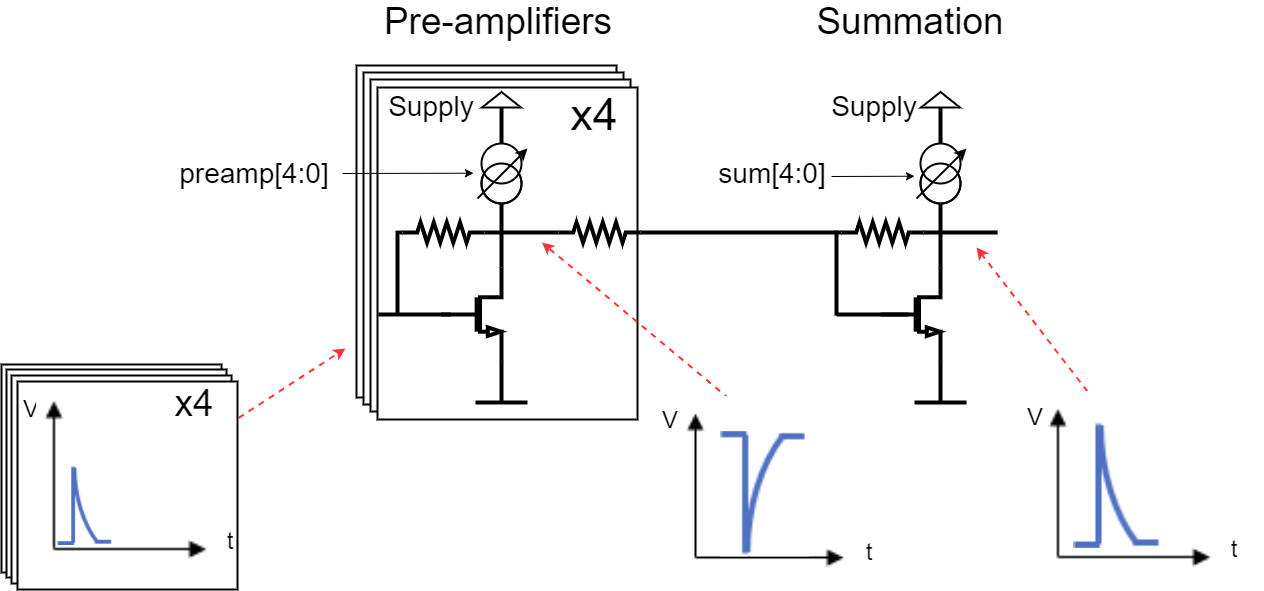}
    \caption{Schematic of the input (left side) and summation (right side) stages. The input stage is made of four pre-amplifiers with individual inputs and one common output. For illustrative purposes, some generic pulse waveforms are shown to display the polarity of the pulses throughout the cascade.}
    \label{fig:tia_sum}
\end{figure}

The pre-amplifiers of the input stage, shown in Fig.~\ref{fig:tia_sum}, work in voltage mode as they amplify the voltage pulse generated across the external resistor by the SiPM and produce an output voltage pulse. They also need to provide proper termination to the transmission lines connecting to the SiPM, which is essential for preserving the pulse shape. 
Inside the camera, these lines can reach a length in the tens of centimeters to cover the distance between the multi-channel ASIC and its pixels.

The amplifier has a common source configuration with resistive feedback, which defines at the same time the gain and the bias point. The choice of a single ended topology over a differential one provides several design advantages:
\begin{itemize}
    \item higher bandwidth and phase margin due to fewer internal nodes and smaller parasitic capacitors
    \item better power efficiency for the same trans-conductance and output current requirements due to a single branch per amplifier and the quadratic current characteristic of the output buffers
    \item higher SNR due lower thermal noise from the amplifying transistors for the same bias current and to fewer noise contributors overall
\end{itemize} 
An N-type transistor with the highest threshold voltage available (HVT-NMOS) is selected as active device in the pre-amplifiers. This choice extends the voltage headroom available to the output pulse by $\sim150\,$mV compared to standard transistors.

\subsection{Summation stage}
The active summation is implemented using a voltage-mode inverting summing amplifier. The output voltages of the four pre-amplifiers are converted into currents through series resistors referenced to the virtual ground of the summing amplifier. The summation is the result of Kirchhoff's current law applied to this virtual ground node. The current is then converted into an output voltage by the feedback resistor. The amplifier employs the same HVT-NMOS common-source architecture as the pre-amplifiers, thus allowing a DC-coupling among the stages. 
The finite bandwidth of the summing amplifier limits the efficacy of the virtual-ground at its input giving rise to a current partition through the series resistors. The resulting signal attenuation is compensated, for the most part, by properly sizing the feedback resistor.  

\subsection{Output buffers}
An output buffer is required to transmit the voltage pulses to an external ADC through a transmission line with a controlled characteristic impedance. 
A single-ended buffer with adjustable DC voltage and a pseudo-differential buffer. Of the eight pixel channels present on the ASIC, four support the differential output and four support the single-ended one. 
Both designs use a two-stage configuration and include two adjustable AC-coupling capacitors used to modify the characteristic of the pulse-shaping filtering described earlier. 

\begin{figure}
    \centering
    \includegraphics[width=0.8\textwidth]{ 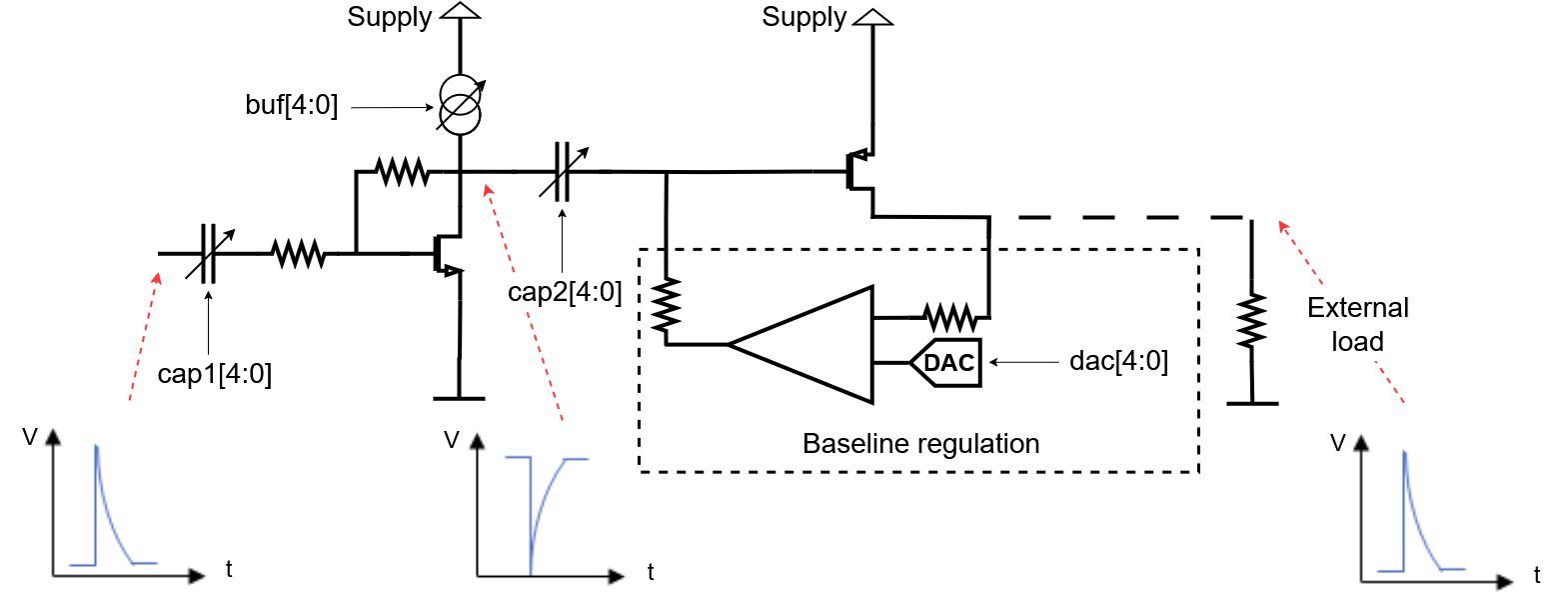}
    \caption{Schematic of the two stages comprising the  single-ended output buffer: amplifier with resistive feedback (left side) and open-loop amplifier with baseline regulation (right side). Generic pulses are added to illustrate the signal polarity at various nodes.}
    \label{fig:se_buff}
\end{figure}

\paragraph{\bf Single-ended topology}
The single-ended topology depicted in Fig. \ref{fig:se_buff} is built out of two AC-coupled inverting stages. The first is an NMOS common-source with resistive feedback and adjustable bias current, whereas the second is a P-type transistor (PMOS) common-source with an external resistive load. The single-ended buffer supports a DC-coupled $50~\Omega$ load connected through a matched transmission line. 
The gain of the second stage is given by the product of the transistor trans-conductance and the external load. 
The second stage uses a low-frequency feedback loop to regulate the output voltage by controlling the gate terminal of the PMOS. This regulation also allows indirect control of the gain as it defines the bias current flowing through the active device.

\begin{figure}
    \centering
    \includegraphics[width=0.8\textwidth]{ 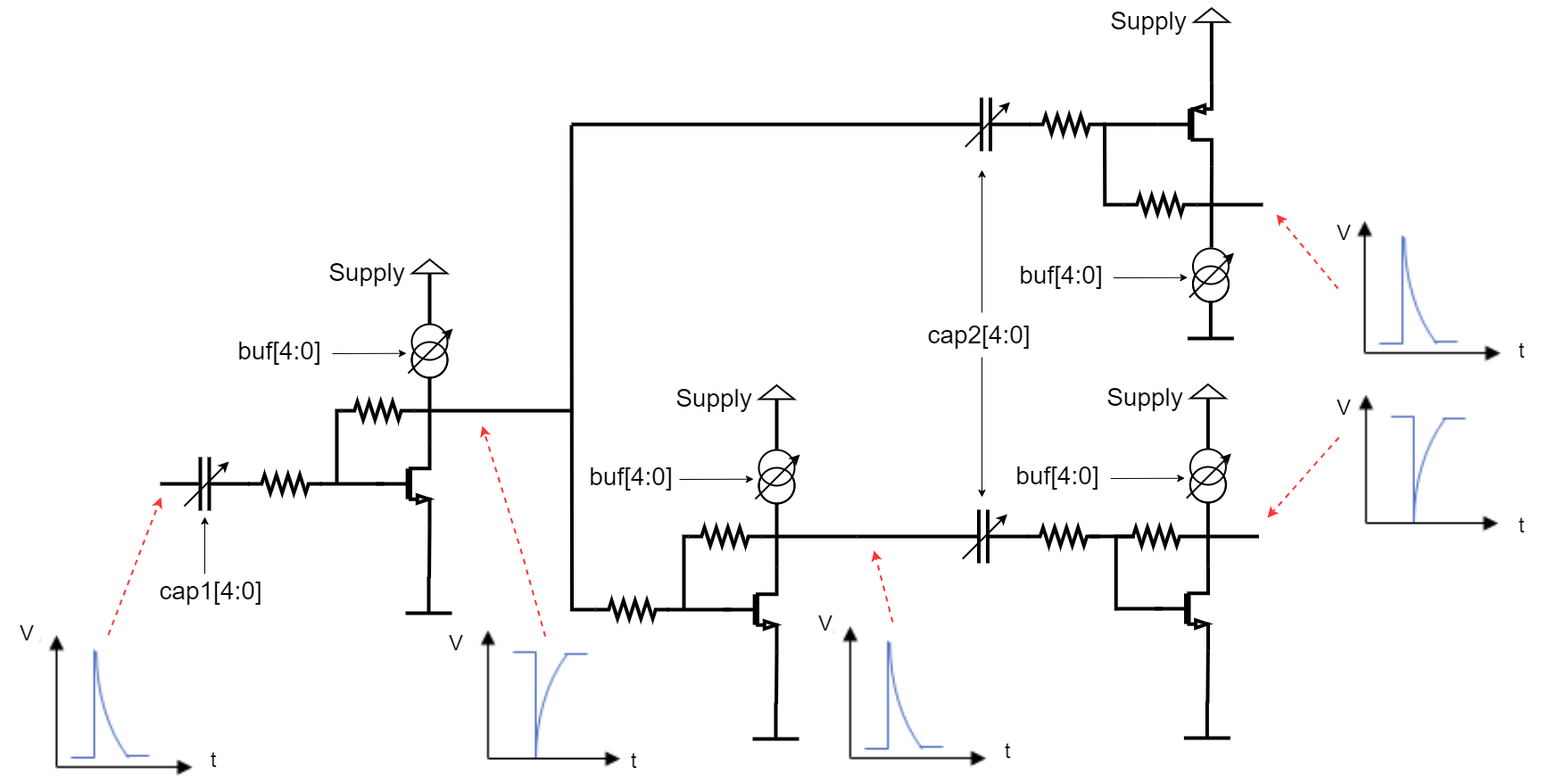}
    \caption{Schematic of the pseudo-differential buffer. The left side and middle amplifiers convert the single-ended input into pseudo-differential. The right-side amplifiers are responsible for driving the external loads. Generic pulses are added to illustrate the signal polarity at various locations .}
    \label{fig:p_diff_buff}
\end{figure}

\paragraph{\bf Pseudo-differential topology}
The pseudo-differential buffer depicted in Fig. \ref{fig:p_diff_buff} also adopts a two-stage topology. The first stage converts the signal from single-ended to  pseudo-differential through a succession of two inverting single-ended amplifiers. The second stage buffers both signals towards an AC-coupled resistive load. The resistive load can be in the form of a single $100\,\Omega$ differential resistance or of two $50\,\Omega$ resistances towards the AC ground. 
All stages are implemented with common source amplifiers with resistive feedback and biased by adjustable current sources. 

\subsection{Pulse-shaping configurations}
\begin{figure}
    \centering
    \includegraphics[width=1\textwidth]{ 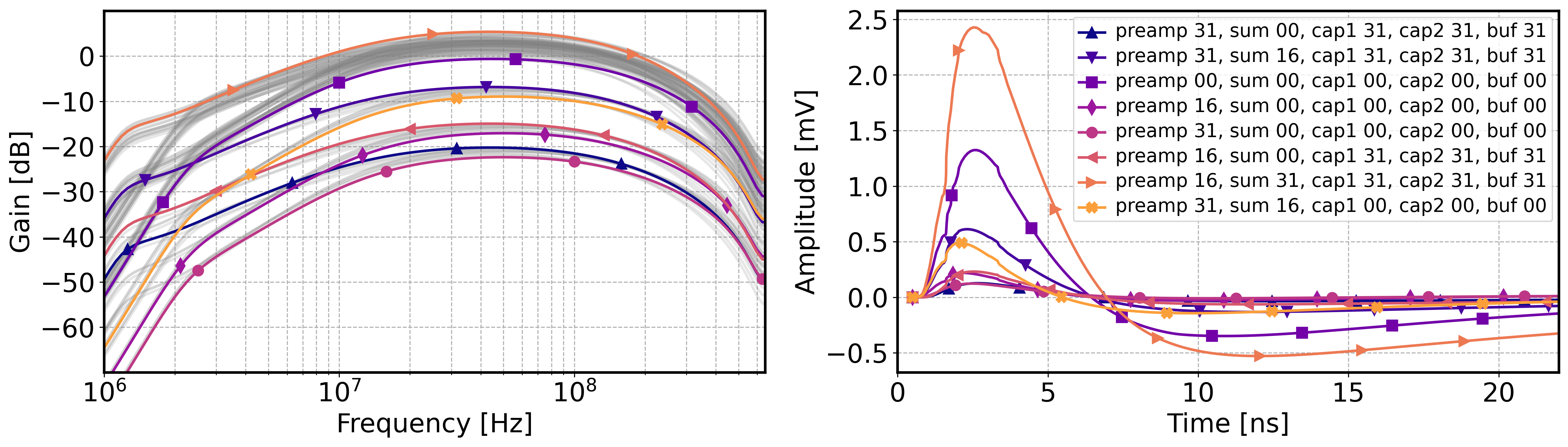}
    \caption{Simulation of the pulse-shaping characteristic of the FANSIC for a sub-set of configurations. Left: Filter characteristics in the frequency domain. The highlighted curves constitute a subset of the most relevant configurations. Right: Simulation of the time domain pulse-shaping induced by the subset on a S13360 SiPM. In both figures the markers are only used for illustration purposes.}
    \label{fig:ac_coupling}
\end{figure}

The band-pass transfer function responsible for shaping the pulse is adjustable by means of digital control words. The high-pass portion is determined by the AC-coupling capacitors along the channels. The value of the external capacitors is defined by the requirement of impedance matching with the transmission lines. 
Conversely, the internal ones (shown in Fig.~\ref{fig:se_buff}, \ref{fig:p_diff_buff}) can be adjusted with two 5-bit words to modify the frequency of the corresponding zeros in the transfer function. 
The low-pass portion is determined by the bandwidth of the channel stages, which depend on the trans-conductance and the parasitic capacitance of each amplifier. The trans-conductances are individually adjustable through a 5-bit control word, which allows to adjust the transfer function gain and the frequency of its poles.
On the whole, the transfer function can be adjusted through 5 parameters, each coded on 5-bits. Of the 3125 possible combinations, only 243 were selected for the simulation of the frequency response. The subset is obtained by combining the minimum, middle and maximum values of each parameter and the simulation results are reported in Fig.~\ref{fig:ac_coupling}.
For the transient response only 9 most representative configurations were selected because of the steep requirement of computational resources.

\section{Experimental results} \label{experimental_results}

\begin{figure}
    \centering
    \includegraphics[width=0.8\textwidth]{ 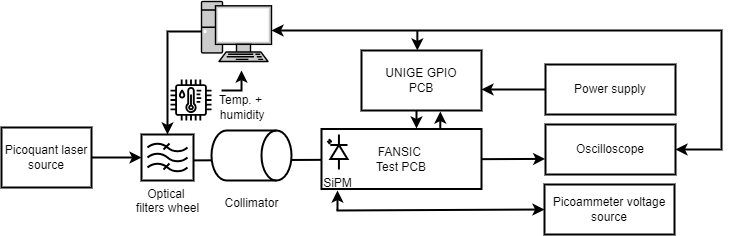}
    \caption{Diagram of the measurement setup for the characterization of the detector comprising FANSIC and the S13360 SiPM.}
    \label{fig:Test_setup}
\end{figure}

The ASIC performance is evaluated together with a $3\times3$\,mm$^2$ S13360 SiPM from Hamamatsu and a pulsed laser source from PicoQuant, which emits light at a wavelength of $375$\,nm. The measurement setup is shown in Fig.~\ref{fig:Test_setup}: the system under test and the optical components are placed in a light-tight metal box, which also shields the electronics from electromagnetic interference. The actuation of the instrumentation has been completely automatized to gather a statistically significant collection of samples. FANSIC output signal is acquired with a digital oscilloscope which employs a sampling rate of $10\,$Gsps. The SiPM is biased at $56\,$V through a picoammeter voltage source resulting in an over-voltage above the breakdown of about $3\,$V. 
Since the light source is only adjustable through a manual potentiometer, the intensity delivered to the sensor's surface is accurately controlled using a set of optical neutral density (ND) filters. These are mounted on an automated rotating wheel from Thorlabs (FW212CNEB) and allow to vary the light intensity delivered to the SiPM beyond $3$ orders of magnitude. A collimator is used to couple the filters to the SiPM while minimizing reflections. 
The UNIGE general purpose inputs outpus (GPIO) board was developed by the University of Geneva to support the characterization of electronics prototypes. 
It is used to provide the required analog, digital and power inputs to the FANSIC test PCB and to allow communication with a PC through a Cyclone $5$ FPGA. 
The temperature and relative humidity are acquired through the Yocto temperature USB sensor from Yoctopuce.

\paragraph{\bf Test PCB}

\begin{figure}
\includegraphics[width=0.55\linewidth]{ 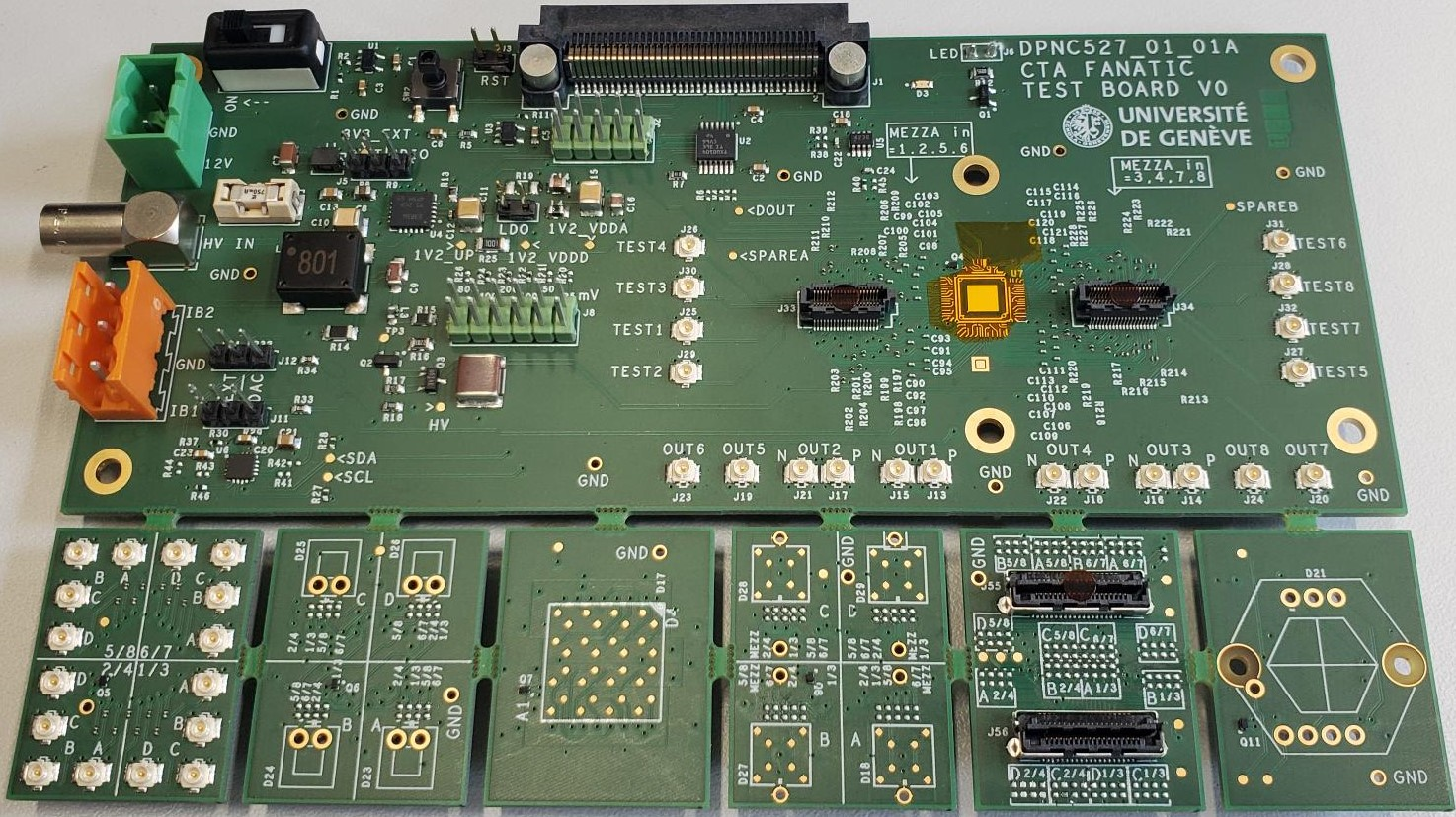}
\hfill
\includegraphics[width=0.3\linewidth]{ 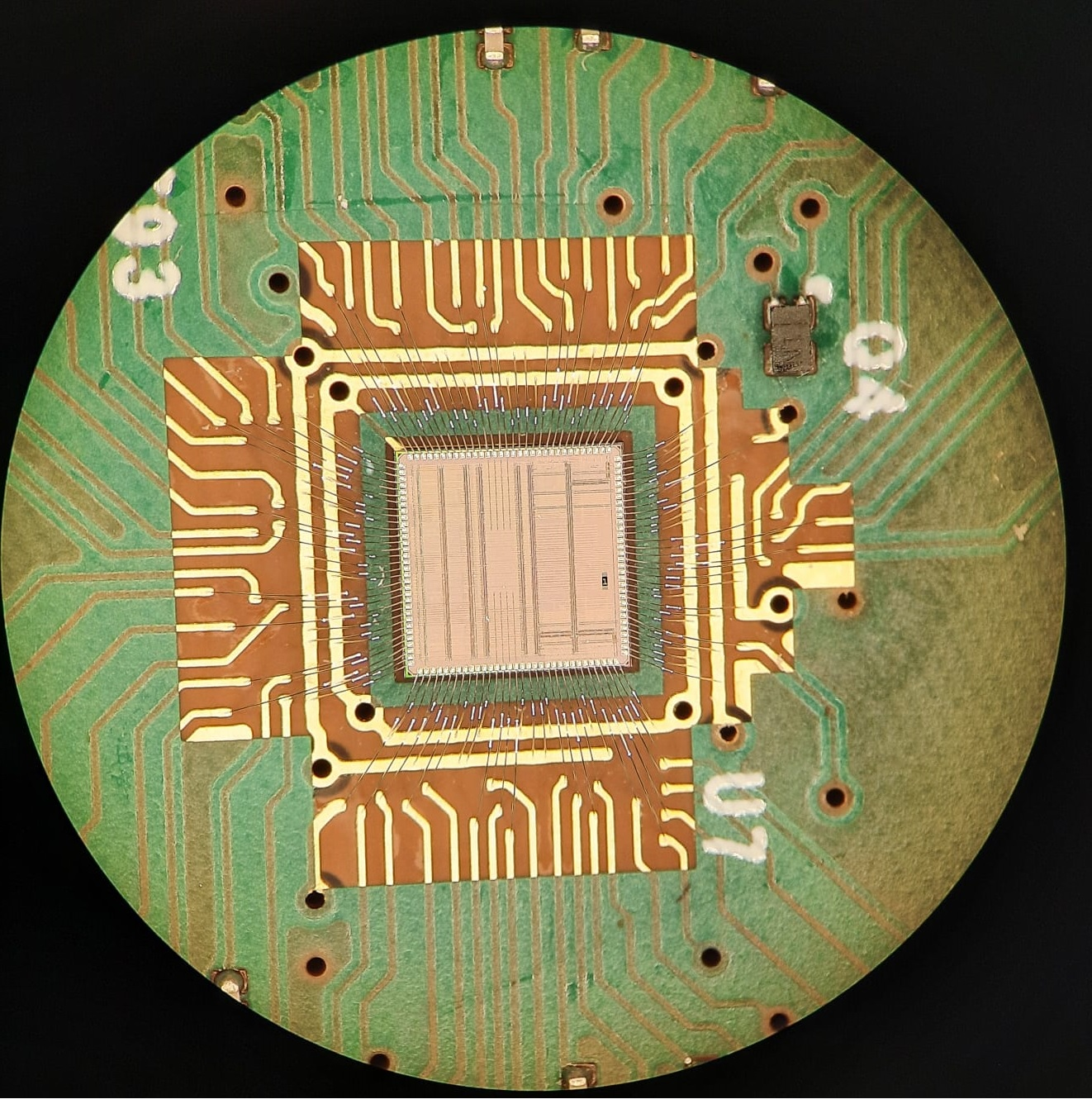}
\caption{Left: Test PCB with mezzanines to connect instrumentation and different SiPMs. Right: Close-up of FANSIC wire-bonded the the PCB with the visible ground and supply rings on the periphery. Die dimensions: $3.586\,\times\,3.586\,\text{mm}^2$.}
\label{fig:PCB}
\end{figure}

The FANSIC chip has been directly wire-bonded to a test PCB (as shown in Fig.~\ref{fig:PCB}) as to minimize the impact of parasitic elements on the validation of the ASIC. 
In addition, a set of mezzanine boards was developed to accommodate different SiPMs and to allow direct connection to an arbitrary waveform generator. The main components that populate the test PCB are:
\begin{itemize}
    \item a low drop-out voltage regulator (LDO) to power FANSIC around $1.2\,$V
    \item a current DAC which provides two reference currents of $90\,\mu$A
    \item a bank of digital level shifters between $1.2$ and $3.3\,$V
    \item dedicated power connectors to directly bias the SiPMs and the FANSIC with external sources.
\end{itemize} 
The LDO and DAC can be configured allowing the evaluation of the ASIC behavior under different operating conditions or can be bypassed altogether via external power connectors. The level shifters provide compatibility with the low voltage CMOS electrical standard which is supported by the majority of commercial FPGAs. The test PCB has been designed to connect directly with the GPIO board.

\subsection{Optical setup characterization } \label{subs:optical_cal}

\begin{figure}
    \centering
    \includegraphics[width=0.8\textwidth]{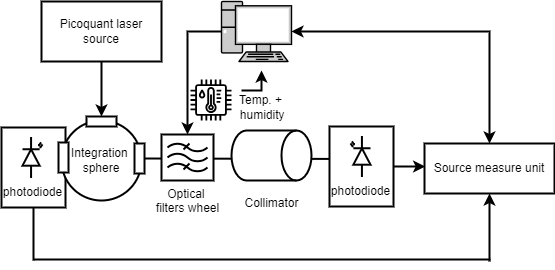}
    \caption{Optical setup diagram for the neutral density filters characterization.}
    \label{fig:Filters_setup}
\end{figure}

The optical setup is characterized by measuring the ratio of the light intensity after the collimator to the value before the filter wheel. This approach provides an expectation for the number of photoelectrons, independent of distortions and saturation introduced by the electronics.
A relative correction is favored over an absolute one, as it eliminates the need to account for variations in PDE, area, and alignment among the different photosensors used.
The setup used for this purpose is shown in Fig.~\ref{fig:Filters_setup}. It was obtained from the one intended for the ASIC characterization by replacing the detector with a simple photodiode (S1337-1010BQ from Hamamatsu) and interposing an integration sphere (IS200-4 from Thorlabs) between the light source and the ND filters. The sphere provides a uniform intensity distribution on its surface, thus enabling measurements of the flux delivered to the filter wheel through an additional photodiode (S1337-1010BQ). The ratio of the two photodiode currents provides a direct measure of the transmittance across the optical path between them. The laser is kept at a fixed intensity setting while the filters are cycled. The two strongest filters are not used because even in combination with the maximum laser intensity the detected number of photoelectrons is not statistically significant. 
A two-channel source measure unit is used to simultaneously acquire, for each filter, ten thousand current samples from the photodiodes. 

\begin{figure}
    \centering
    \includegraphics[width=1\textwidth]{ 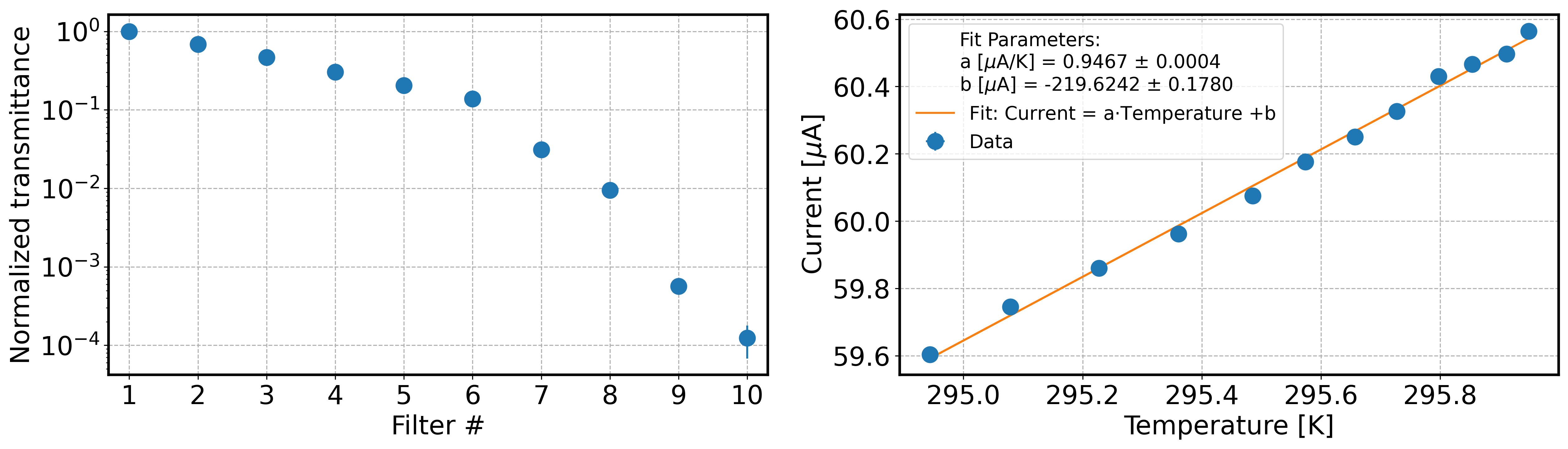}
    \caption{Measured data for the characterization  of the optical setup. Left: Transmittance of the optical path between the two photodiodes as a function of the wheel position. The amplitude is normalized to the position that corresponds to an opening without any filter. Right: absolute current of the upstream photodiode as a function of the temperature. The legend contains the fit parameters and their uncertainty.}
    \label{fig:Filters_meas}
\end{figure}

The measured data is reported in Fig.~\ref{fig:Filters_meas}. On the left, the data points correspond to the ratio of the photodiodes currents normalized to the highest value, which is obtained with the first wheel position. On the right, the plot of the upstream photodiode current versus temperature confirms the expected dependence of circa $+15\%$ per degree. On the other hand, the transmittance measurements only show a dependence around $1\,$ppm per degree, being the ratio of two currents with the same temperature dependence.  

\subsection{ASIC characterization}
\label{sec:asic_charac}

FANSIC was characterized with the setup for the detector system previously described. The channel transfer function was configured with the parameters detailed in Tab.~\ref{tab:dig_conf}. Such configuration provides optimal performance when paired with a S13360 SiPM, such as single-photon resolution. 
The data from \ref{subs:optical_cal} is used to estimate the average number of photoelectrons corresponding to different filters. 

\begin{table}
\centering
\resizebox{\textwidth}{!}{\begin{tabular}{lcccccc}
\toprule
\midrule
Control word & preamp[$4:0$] & sum[$4:0$] & cap1[$4:0$] & buf[$4:0$] & cap2[$4:0$] & dac[$4:0$] \\
\midrule
Decimal value  & $16$         & $31$        & $4$       & $8$           & $8$ & $0$ \\
Electrical value & $2\,$mA    & $3.3\,$mA   & $22\,$pF  & $2\,$mA       & $36.5\,$pF & $20.5\,$mV \\ 
\midrule
\bottomrule
\end{tabular}}
\caption{Optimal FANSIC configuration which provides the desired performance when coupled with a S13360 SiPM. The electrical components controlled by the digital words are depicted in Fig. \ref{fig:tia_sum} and \ref{fig:se_buff}.}
\label{tab:dig_conf}
\end{table}

\paragraph{\bf Output pulse features}

\begin{figure}
    \centering
    \includegraphics[width=0.9\textwidth]{ 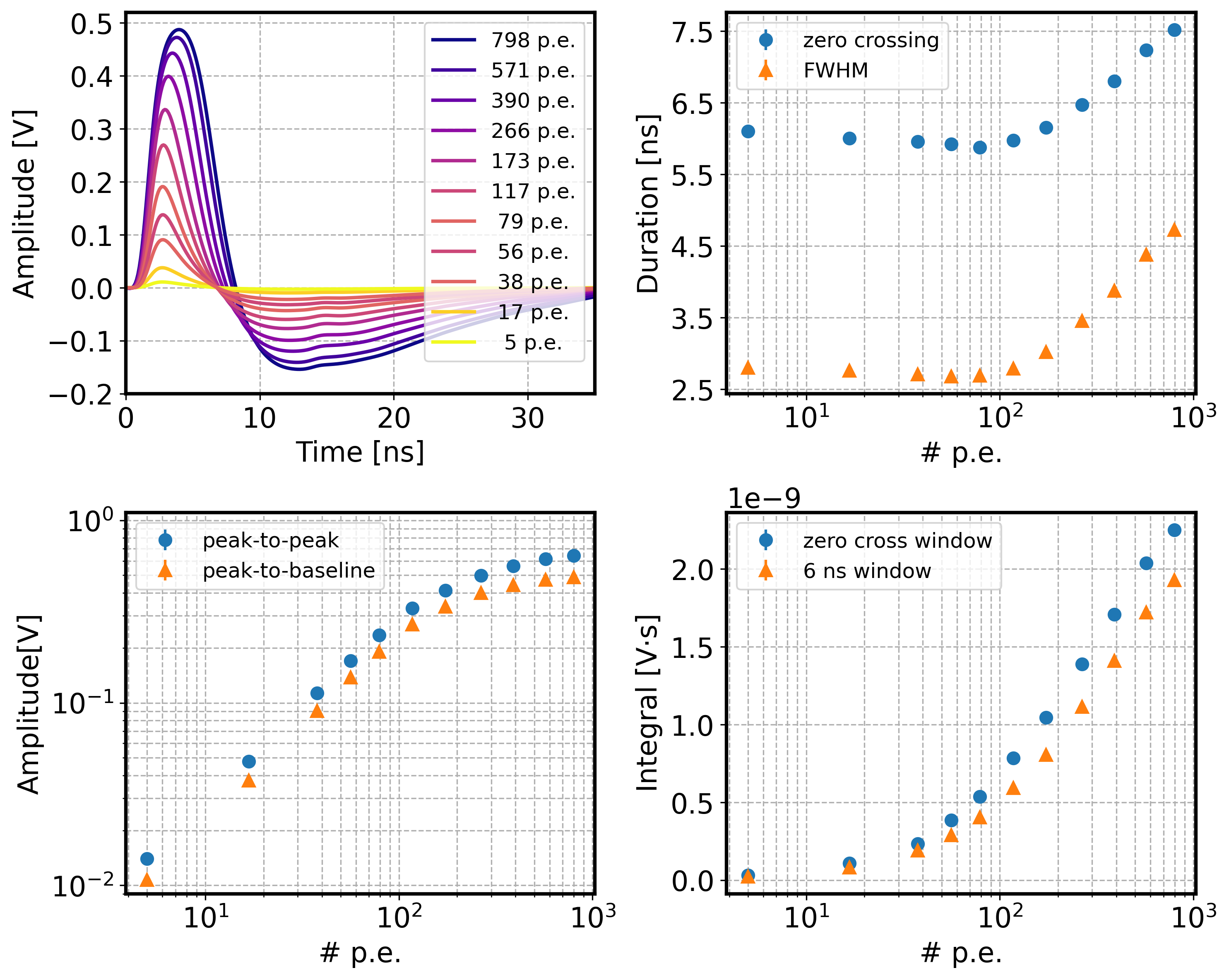}
    \caption{Measurement of the output pulses features for different light intensities. Top left: average waveforms from ten thousand samples for each intensity. Top right: zero-crossing and FWHM durations versus the number of photoelectrons. Bottom left: pulse peak-to-peak and peak-to-baseline amplitudes versus the number of photoelectrons. Bottom right: area under the pulses with the zero-crossing and $6\,$ns integration windows.}
    \label{fig:ND_pulses}
\end{figure}

The output pulse features are evaluated over a wide range of light intensities spanning from $0$ to $\sim800\,$p.e., extending well beyond the expected operating conditions. The acquired pulses and their features are reported in Fig.~\ref{fig:ND_pulses}. 

As the light intensity varies, various regimes can be identified in the response of the frontend. In the lower range, up to $\sim20\,$p.e., the output pulses maintain a consistent shape with fairly constant duration and a distinguishable correlation between the peak amplitude and the number of photoelectrons. A clear separation was observed by stacking waveform samples synchronously using the trigger signal of the light source.

At medium intensities, up to $\sim100\,$p.e., the deviation of the pulse characteristics from the expected values is still negligible. Nevertheless, it is not possible to use the peak amplitude to clearly distinguish different numbers of photoelectrons due to the larger noise contributions. Therefore, the area under the fast portion of the pulse is used as an indicator for the number of photoeletrons. This approach provides better correlation and removes part of the uncorrelated noise by averaging its contribution over the integration window. A fixed integration window of $6$\,ns is used for the frontend characterization because it captures the largest part of the positive pulse area in the region of interest (up to $250\,$p.e.). In addition, this approach is compatible with the digital charge reconstruction that will be implemented in the camera using a fixed sampling rate of $1\,$GHz.

At higher intensities, up to $\sim800\,$p.e., the electronics approaches a saturation regime where the peak amplitude reaches a plateau (of about $500\,$mV) and the pulse durations increase with the number of photoelectrons, resulting in distorted waveforms. 
Circuit simulations show that the first frontend components to reach saturation are the pre-amplifiers within the input stage. 

For intensities above $800\,$p.e., the pulse peaks exceed the supply voltage of FANSIC triggering the activation of the input over-voltage protections. These circuits clip the voltage to safe values and provide an additional path towards ground to the avalanche charges. The impedance discontinuity induces considerable reflections along the transmission line, which heavily affect the shape of the output pulse. The resulting distortion, together with the charge loss through the protection circuits, 
renders the pulse inapt for the quantification of photoelectrons but still allows the identification of excessive illumination.

\begin{figure}
    \centering
    \includegraphics[width=0.8\textwidth]{ 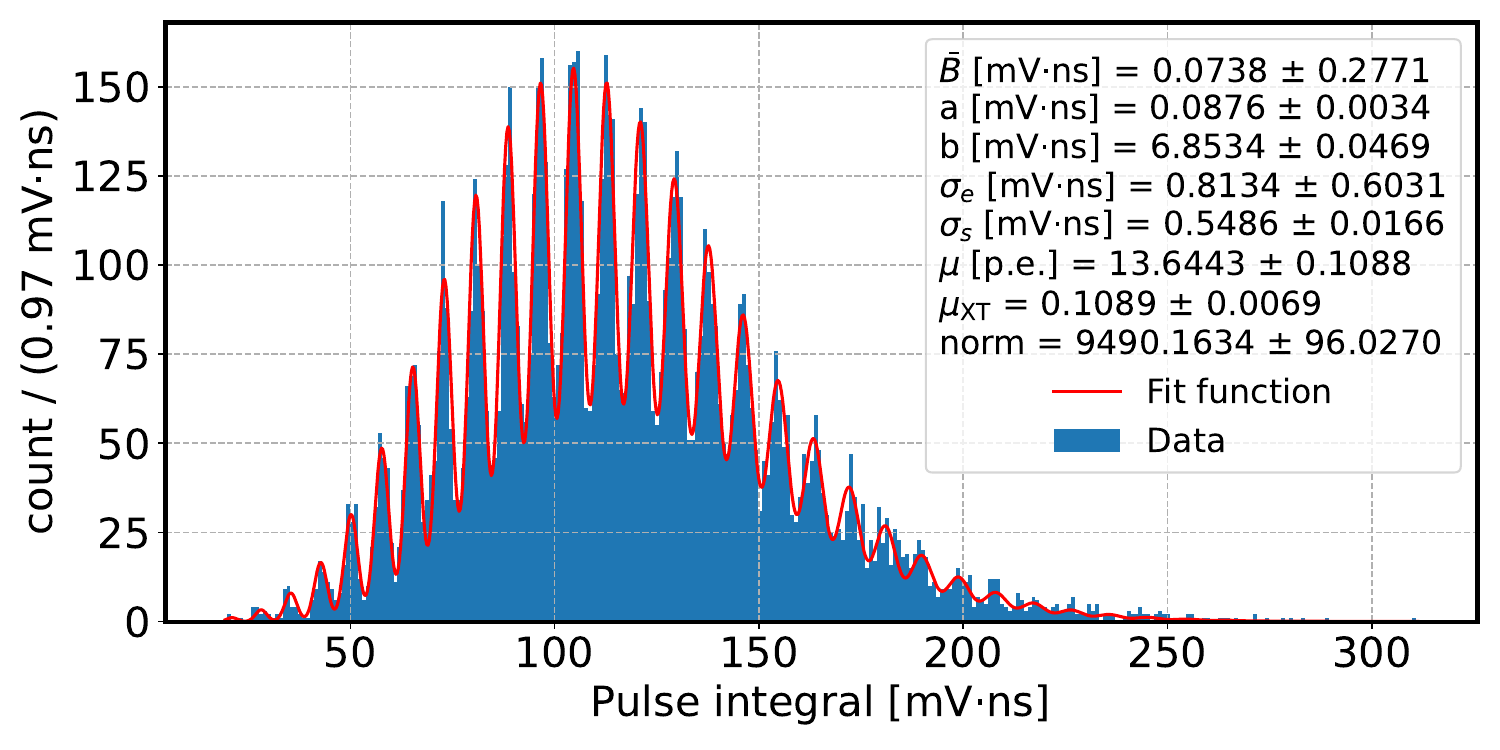}
    \caption{The multi-photoelectron distribution was measured under a fixed light intensity, corresponding to an mean value of $\sim13\,$p.e. The vertical axis shows the density of occurrences per bin, while the horizontal axis represents the integral of the output pulses over a 6\,ns time window. The red curve illustrates the fit using a generalized Poisson pdf. The legend includes the fit parameter values along with their associated uncertainties.}
    \label{fig:MPEs}
\end{figure}

\paragraph{\bf Multi photoelectron response}
The multi photoelectron response is used to evaluate the SNR and linearity performance of the detector system, which comprises the frontend and the SiPM. The evaluation procedure involves fitting a histogram of the pulse integrals with a generalized Poisson probability density function (pdf), as described in~\cite{Alispach:2020} (equation 3.2). This method relies on a clear separation between the histogram's peaks, therefore its applicability is limited to the lower illumination intensities. A fitted histogram containing occurrences up to $\sim30\,$p.e. is shown in Fig.~\ref{fig:MPEs}.

\begin{figure}
    \centering
    \includegraphics[width=0.7\textwidth]{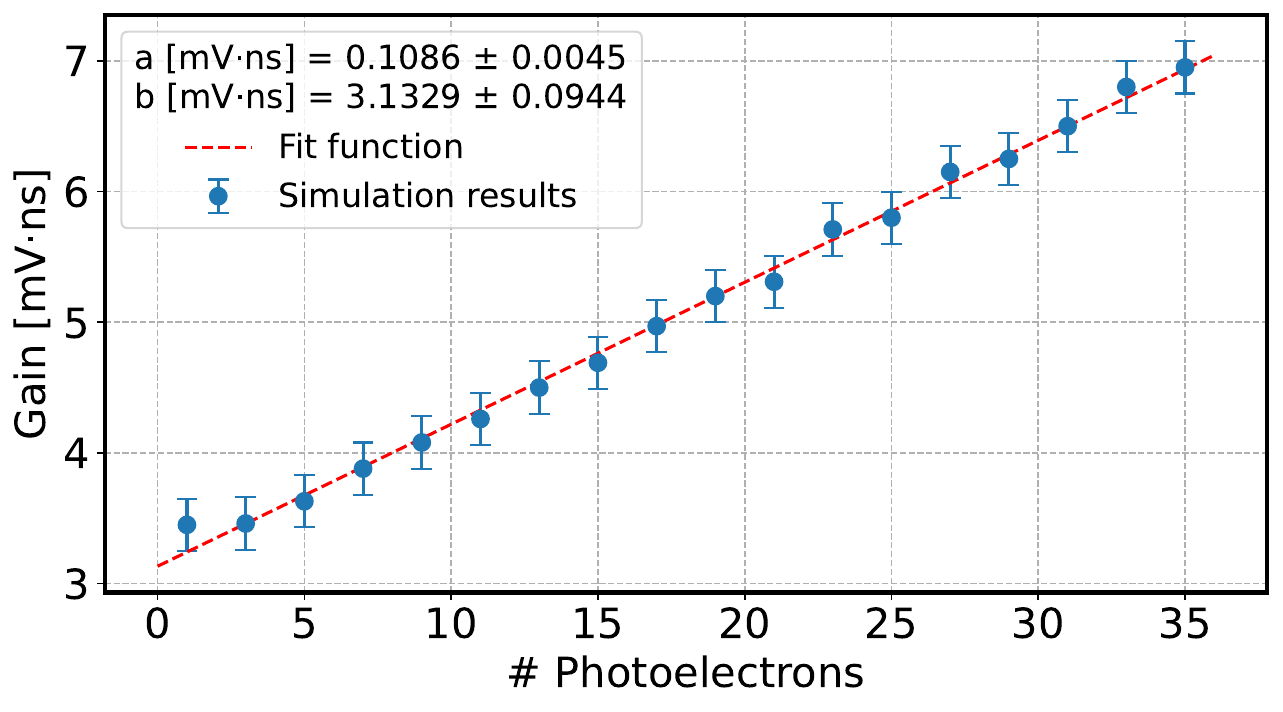}
    \caption{Circuit simulation of the gain $G(N_\mathrm{pe})$ as a function of the number of photoelectrons ranging from $1$ to $35\,$p.e. The gain represents the increase in pulse area per photoelectron, where the area is calculated using the $6\,$ns integration window.}
    \label{fig:gain_sim}
\end{figure}

The gain is defined as the increment of the pulse integral per photoelectron and, therefore, corresponds to the separation between adjacent MPE peaks.  
As predicted by simulations (Fig.~\ref{fig:gain_sim}) and confirmed by measurements (Fig.~\ref{fig:ND_pulses} bottom right), at lower intensities the pulse area has a quadratic dependence on the number of photoelectrons. The deviation from a constant behavior is caused by the quadratic dependence of the amplifying transistors' current from their controlling voltage. This implies that the gain, which is a crucial parameter of the generalized Poisson used to fit the measurements, has to be modeled with a linear function such as:

\begin{equation}
    G(N_\mathrm{pe}) = (N_\mathrm{pe}-1) \cdot a + b  
\label{eq:gain}
\end{equation} 
where the linear $a$ and constant $b$ coefficients are estimated by the fit of the MPE histogram. 
The discrepancy between the coefficient values obtained from simulations and measurements is attributed to inaccurate modeling of the parasitic elements present in the actual measurement setup.

The other parameters evaluated through the fit are: the distribution baseline ($\bar{B}$) that corresponds to $0\,$p.e., the electronics ($\sigma_\mathrm{e}$) and sensor ($\sigma_\mathrm{s}$) contributions to standard deviation of the histogram peaks, the mean photoelectron number ($\mu$) of the Poisson distribution, the optical cross-talk probability per photoelectron ($\mu_\mathrm{XT}$) and the norm that represents the size of the population used for the computation. The most relevant fit parameters are summarized in Tab.~\ref{tab:fit_params}.

Each MPE peak is a Gaussian distribution whose mean value represents an integer number of photoelectrons. The envelope of the MPE distribution is an exponentially modified Gaussian function centered around the mean value of the pulse area for that particular dataset, namely $\mu\cdot G(\mu)$.

\begin{table}
\centering
\begin{tabular}{cccccc}
\toprule
\midrule
 & \multicolumn{2}{c}{Gain} & $\sigma_e$ & $\sigma_s$ & $\mu_{XT}$ \\
 & a & b &   &   &   \\
 & [mV$\cdot$ns] & [mV$\cdot$ns] & [mV$\cdot$ns] & [mV$\cdot$ns] & [\%] \\
Value & 0.0876 & 6.8534 & 0.8134 & 0.5486 & 10.9 \\
Fit uncertainty & 0.0033 & 0.0456 & 0.03 & 0.0068 & 0.68 \\ 
\midrule
\bottomrule
\end{tabular}
\caption{Main parameters of the MPE fit of Fig.~\ref{fig:MPEs}. The uncertainties are the entries of the covariance matrix produced by the fit.}
\label{tab:fit_params}
\end{table}

\paragraph{\bf Signal to Noise ratio}
The overall SNR takes into account both contributions of the electronics and the SiPM, its dependence on the number of photoelectrons and is detailed by the equation:
\begin{equation}
    \mathit{SNR}(N_\mathrm{pe}) = \frac{G(N_\mathrm{pe})}{\sqrt{\sigma_e^2 + N_\mathrm{pe}\sigma_s^2}}
    \label{eq:snr_equation}
\end{equation}
This implies that as $N_\mathrm{pe}$ increases the SNR degrades, as confirmed by the growing spread of the histogram's peaks. The SNR for $1\,$p.e. is the value relevant for false positives and missing event detections in the trigger. Using the fit parameters from Tab.~\ref{tab:fit_params} yields a value of $6.98\pm0.22$. Whereas, at the threshold of the single-photon detectability, approximately $33\,$p.e., the SNR drops below $3$.

Ignoring the sensor contribution gives the intrinsic SNR of the electronics, previously defined in Eq.~\ref{eq:snr}, whose estimated value for $1\,$p.e. is $8.42\pm0.31$. This means that the single-photon resolution range can be extended by employing a photodetector that shows a better noise performance.

\begin{figure}
     \centering
     \includegraphics[width=0.7\linewidth]{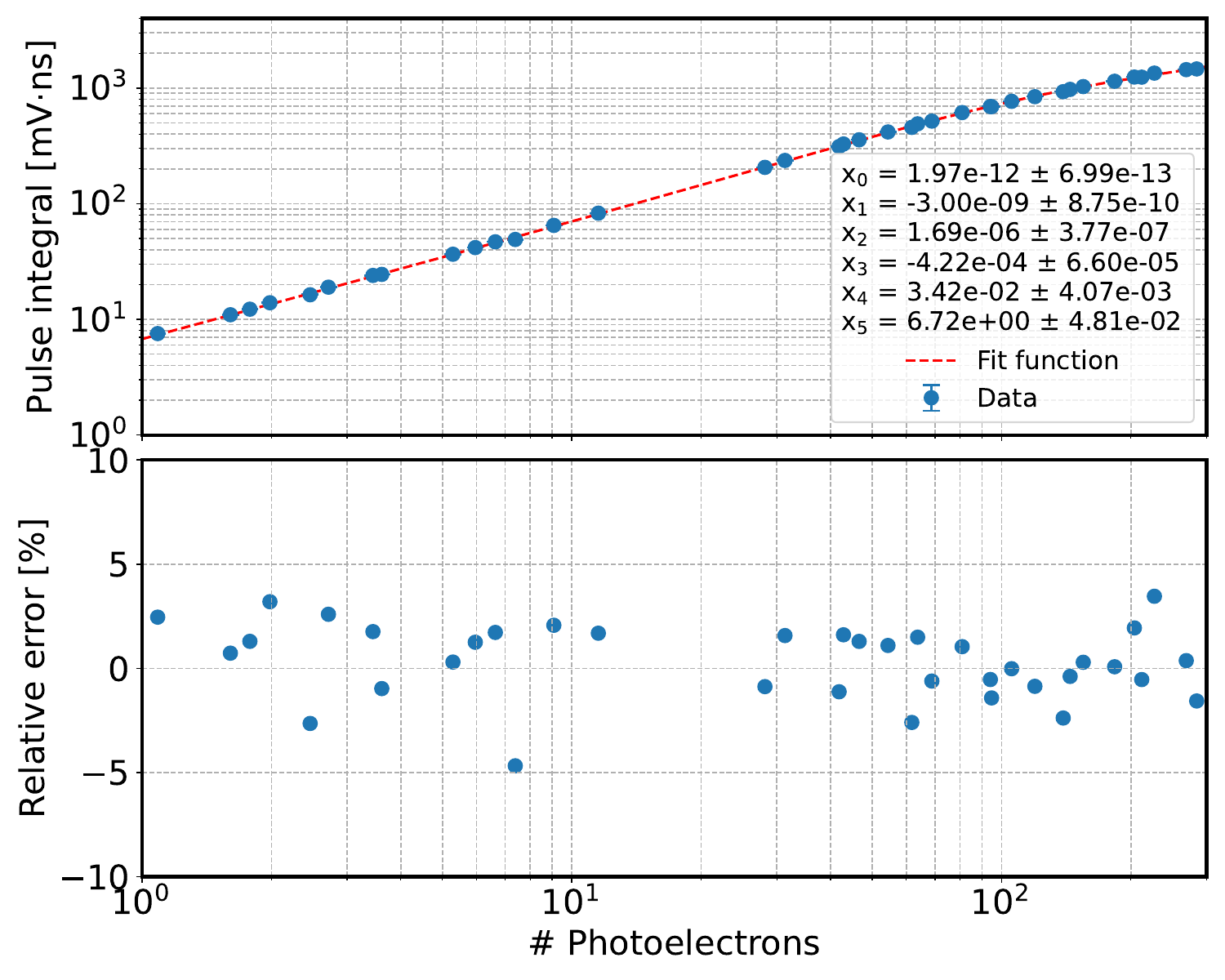}
     \caption{The transfer characteristic, mapping the number of photoelectrons to the integral of the ASIC output pulse, was measured across the dynamic range. Top: the measurement data is shown alongside an interpolation using a sixth-degree polynomial. The legend includes the fit coefficients and their uncertainties. Bottom: plot of the relative error, in percentage, of the measurement points with respect to the interpolation.}
     \label{fig:linearity}
 \end{figure}

\paragraph{\bf Linearity}
The transfer function from number of photoelectrons to pulse integrals is obtained by combining the laser intensity settings with the ND filters. This provides a reasonable variety of data points across the entire dynamic range, as shown in Fig.~\ref{fig:linearity}. Since the fit only works in the lower part of the dynamic range, the number of photoelectrons is extrapolated using the relative transmittance described in subsection~\ref{subs:optical_cal}. For each intensity setting, the data corresponding to one of the highest attenuation filters is used to extract the mean number of photoelectrons (i.e. the $\mu$ of the fit). 
Then when the fit is not applicable because of excessive sensor noise, the number of photoelectrons is estimated by correcting a previous measurement for the calibrated transmittance of the filters. 

The pulse integral values displayed on the vertical axis correspond to the mean of a multi-sample measurement obtained with fixed intensity and ND filter, and corrected for the cross-talk according to:

\begin{equation}
    \bar{y}(N_\mathrm{pe}) = (1-\mu_\mathrm{xt}) \cdot \frac{1}{n_\mathrm{meas}}\sum_{i=1}^{n_\mathrm{meas}} \int_{t_0}^{t_1}  V_\mathrm{i}(t) \cdot dt\
    \label{eq:charge_npe}
\end{equation}
where $n_\mathrm{meas}$ is the number of samples used to calculate the average, $V_\mathrm{i}(t)$ is the i-th pulse at the ASIC output, $t_0$ and $t_1$ are the zero-crossing times defining the integration window.

The calibration of the transfer function over the entire dynamic range is done with a sixth-degree polynomial, whose coefficients are reported in the legend of Fig.~\ref{fig:linearity}. After calibration, all measurement points with the exception of a single outlier show a deviation from the model within the required $5\%$.

\section{Conclusions} \label{conclusions}
This work presents FANSIC, a high-performance analog front-end ASIC designed for the fast and low-noise readout of large-area SiPMs suitable for cameras in IACTs. Fabricated in 65\,nm CMOS technology, FANSIC is optimized to meet the stringent requirements of next-generation gamma-ray observatories, particularly in terms of single-photon resolution, fast response times, and low power consumption.

Experimental results demonstrate that FANSIC achieves a pulse FWHM of $3\,$ns across a dynamic range of $1\text{--}250\,$p.e., enabling rapid signal processing of Cherenkov light from events with energies as low as 20\,GeV. The ASIC provides single-photoelectron resolution up to 30\,p.e., which is critical for detecting faint events, and maintains a post-calibration non-linearity around $5\%$ up to $300\,$p.e. With a power consumption of only 23\,mW per pixel from a 1.2\,V supply, FANSIC is an energy-efficient solution for large-scale cameras requiring fine spatial resolution. The dynamic range extends to up to $800\,$p.e., albeit with some degradation in its performance metrics such as FWHM and resolution.

FANSIC incorporates active summation stages and configurable signal processing elements to address signal pile-up from night-sky background photons and adapt its pulse-shaping characteristic to different SiPMs.
Its support for both single-ended and pseudo-differential outputs ensures compatibility with a variety of digitization solutions. A future version of the chip will include a configurable output switch to further enhance integration flexibility.

The successful characterization of FANSIC, through both simulations and laboratory measurements, highlights its potential as a key enabling technology for future Cherenkov telescope cameras. This work lays the foundation for the development of scalable, high-performance readout electronics tailored to the evolving needs of astrophysical and particle physics experiments.

\section{Acknowledgments}
Supported by the Swiss National Foundation FLARE grant 20FL21 201539. 

\bibliographystyle{elsarticle-num-names}
\typeout{}

\end{document}